\documentclass[aps,prb,twocolumn,nofootinbib,superscriptaddress]{revtex4-1}

	\usepackage{amsmath, amssymb, graphicx}
\usepackage[section]{placeins}
\usepackage{graphicx}% Include figure files
\usepackage{dcolumn}% Align table columns on decimal point
\usepackage{bm}% bold math

\newcommand{\bra}[1]{\langle #1 \rvert}

\newcommand{\ket}[1]{\rvert #1 \rangle}

\newcommand{\e}{\mathrm{e}}

\newcommand{\be}{\begin{equation}}
\newcommand{\ee}{\end{equation}}
\newcommand{\beq}{\begin{eqnarray}}
\newcommand{\eeq}{\end{eqnarray}}
\newcommand{\nn}{\nonumber}

\def\H1{\widehat{H}_1}

\newcommand{\pd}{\partial}

\def\op#1{{\Hat{\mathrm{#1}}}}

\def\bra#1{\ensuremath{\langle{#1}\vert}}
\def\ket#1{\ensuremath{\vert{#1}\rangle}}
\def\bracket#1#2{\ensuremath{%
    \langle{#1}\mkern1.2mu\vert\mkern1.2mu{#2}\rangle}}

\newcommand{\lb}{\left[}
\newcommand{\rb}{\right]}
\newcommand{\lp}{\left(}
\newcommand{\rp}{\right)}

\begin{document}

% Use the \preprint command to place your local institutional report
% number in the upper righthand corner of the title page in preprint mode.
% Multiple \preprint commands are allowed.
% Use the 'preprintnumbers' class option to override journal defaults
% to display numbers if necessary
%\preprint{}

%Title of paper
\title{Enabling Adiabatic Passages between Disjoint Regions in Parameter Space through Topological Transitions}

% repeat the \author .. \affiliation  etc. as needed
% \email, \thanks, \homepage, \altaffiliation all apply to the current
% author. Explanatory text should go in the []'s, actual e-mail
% address or url should go in the {}'s for \email and \homepage.
% Please use the appropriate macro foreach each type of information

% \affiliation command applies to all authors since the last
% \affiliation command. The \affiliation command should follow the
% other information
% \affiliation can be followed by \email, \homepage, \thanks as well.
\author{Tiago Souza}
\email[]{tsouza@bu.edu}
\author{Michael Tomka}
\affiliation{Department of Physics, Boston University, 590 Commonwealth Ave., Boston, MA 02215, USA}
\author{Michael Kolodrubetz}
\affiliation{Department of Physics, Boston University, 590 Commonwealth Ave., Boston, MA 02215, USA}
\affiliation{Department of Physics, University of California, Berkeley, CA 94720, USA}
\affiliation{Materials Sciences Division, Lawrence Berkeley National Laboratory, Berkeley, CA 94720, USA}
\author{Steven Rosenberg}
\affiliation{Department of Mathematics and Statistics, Boston University, 111 Cummington Mall, Boston, MA 02215, USA}
\author{Anatoli Polkovnikov}
\affiliation{Department of Physics, Boston University, 590 Commonwealth Ave., Boston, MA 02215, USA}

\begin{abstract}
We explore topological transitions in parameter space in order to
enable adiabatic passages between regions adiabatically disconnected
within a given parameter manifold.
To this end, we study the Hamiltonian of two coupled qubits
interacting with external magnetic fields, and make use of the analogy
between the Berry curvature and magnetic fields in parameter space,
with spectrum degeneracies associated to magnetic charges.
Symmetry-breaking terms induce sharp topological transitions on these
charge distributions, and we show how one can exploit this effect to
bypass crossing degeneracies.
We also investigate the curl of the Berry curvature, an interesting
but as of yet not fully explored object, which together with its
divergence uniquely defines this field.
Finally, we suggest a simple method for measuring the Berry curvature,
thereby showing how one can experimentally verify our results.
\end{abstract}

% insert suggested PACS numbers in braces on next line

%\maketitle must follow title, authors, abstract, \pacs, and \keywords
\maketitle

% body of paper here - Use proper section commands
% References should be done using the \cite, \ref, and \label commands

\label{Section: Introduction}
\section{Introduction}
Geometric ideas always played an important
role in the understanding and unification of physical phenomena, the
most prominent example demonstrating this synergy being general
relativity.
More recently, the discovery of topological
insulators~\cite{2005PhRvL..95n6802K,Bernevig:2006ij,Konig.Molenkamp.2007Science.318,Hasan2010}
brought a huge interest in the subject of topology to the field of
condensed matter physics.
The manifestation of geometry in quantum systems evolving adiabatically
was first described by M.\ V.\ Berry~\cite{Berry1984} in 1984.
In this seminal paper, he showed the existence of a phase with the
remarkable geometric property of depending only on the path taken in
parameter space and not on the rate of evolution. This geometric phase
is nowadays known as the Berry phase.

We consider the case where the Hamiltonian of a system
$H(\vec{\lambda})$ depends on three real-valued parameters
$\vec{\lambda}=(\lambda^{1}, \lambda^{2}, \lambda^{3})^{T} \in
\mathbb{R}^3$, thereby describing a three-dimensional parameter
space.
Focusing on the ground-state manifold, the Berry phase $\gamma$
acquired by $\ket{\Psi_{0}(\vec{\lambda})}$ after the parameters
evolve adiabatically along a closed path $C$ reads
\be
\label{eq:geomphase}
\gamma(C)
=
\oint_{C} \vec{A} \cdot d\vec{\lambda}
=
\iint_{S} \vec{F} \cdot d\vec{S},
\ee
where $\vec{A} = i \bra{\Psi_{0}} \vec{\nabla} \ket{\Psi_{0}}$ is the
Berry connection.
The last equality defines the Berry curvature
$\vec{F} = \vec{\nabla} \times \vec{A}$, where the surface $S$ is
bounded by the path $C$.
The Berry connection behaves like a $U(1)$ gauge potential and therefore
cannot directly be observed, whereas the Berry curvature is a local
and gauge-invariant object manifesting the geometric properties of its
associated eigenstate.

An analogy with electromagnetism (E\&M), also presented by
M.\ V.\ Berry~\cite{Berry1984}, shows that the Berry connection plays
the role of a magnetic vector potential and yields through its curl
the Berry curvature, which can be interpreted as an effective magnetic
field.
For each degeneracy in the spectrum, one can choose a closed Gaussian surface $\Sigma_{i}$ that encloses it in parameter space.
The flux of the Berry curvature through $\Sigma_{i}$ defines a
topological quantized invariant
\be
\label{eq:chnum}
\mathrm{ch}_{1}
=
\frac{1}{2\pi} \iint_{\Sigma_{i}} \vec{F} \cdot d\vec{\Sigma}_{i},
\ee
known as the first Chern number.
By noting that
$\vec{\nabla} \cdot \vec{F} = \vec{\nabla} \cdot (\vec{\nabla} \times \vec{A})$, 
one can see that the Berry curvature has zero divergence except
at singularities. These singularities correspond to the degeneracies
in the spectrum of the Hamiltonian, which play the role of effective
magnetic charges in parameter space.
The first Chern number quantization simply reflects the quantization
of these magnetic charges.
Various systems illustrating this analogy have been studied, each
exhibiting different monopole charge configurations in parameter
space~\cite{Oh2009, Li2014, Wu2011, Wiemer2004, Sjoqvist2010, Viennot2006, Nesterov2008, Bruno2006}.

In this paper, we study a system of two coupled qubits which exhibits
sharp topological transitions from continuous closed surfaces carrying
a magnetic charge density to discrete magnetic charges in parameter
space. We then show how introducing symmetry-breaking terms to the
Hamiltonian, one can bypass these closed degeneracy surfaces and open
adiabatic passages between topologically disjoint regions. This
illustrates how one can make adiabatic transitions between different
topological magnetic charge configurations. Such method allows access
to the entire parameter space, and might facilitate the engineering of
entangled states for quantum computation and quantum
information~\cite{quantumcomputationandinformation}.

In addition, we address the issue of the previously presented analogy
with E\&M  not being complete, since in general the Berry curvature
generated by more than one degeneracy is not the same as the
superposition of the effective magnetic fields of individual Berry
monopoles situated at the degeneracies. The superposition principle is
then not necessarily obeyed. We also present scenarios where
degeneracies show a vanishing Chern number, which is equivalent to a
zero effective magnetic charge. In such cases, the curl of the Berry
curvature is shown to be non-vanishing, and thus serves as a probe to
identify such points experimentally. Finally, new sources akin to
electric currents appear alongside the well-known magnetic charges in
the curl of $\vec{F}$.

The paper is organized as follows. In Section~\ref{Section: Two Qubit System} we introduce the interacting system investigated, and analyze two different scenarios in Sections~\ref{Section: Heisenberg interaction} and \ref{Section:XXZ interaction}, where the degeneracies of the system create disjoint regions not adiabatically connected in the parameter space. Then, in Section~\ref{Section: Bypassing degeneracy crossings}, we outline how one can gain access to those forbidden regions by adiabatically breaking and reintroducing symmetries in the Hamiltonian, with topological arguments ensuring that such procedure is robust. We further discuss the symmetry-broken cases in Sections~\ref{Section: Anisotropic fields} and \ref{Section: Broken exchange symmetry}. Finally, we present analytical and numerical analysis of a behavior repeatedly observed for all the studied cases on the curl of the Berry curvature in Section~\ref{Section: Curl analysis}, and present our conclusions in Section~\ref{Section: Conclusion}. More detailed analytical calculations can be found in the Appendix Sections.

% Put \label in argument of \section for cross-referencing
%\section{\label{}}

\section{Two Qubit System}
\label{Section: Two Qubit System}
We consider a system of two interacting
qubits (represented here by quantum spins$-1/2$), coupled to tunable
external magnetic fields.
This choice was inspired by a recent experiment which measured the
Berry curvature~\cite{Roushan2014}.
The Hamiltonian of the system is given by
\be{\label{eq:orighamiltonian}}
H
=
\vec{B} \cdot \left(\gamma_1 \vec{\sigma}_1 + \gamma_2 \vec{\sigma}_2\right)
+
\frac{g}{2}(\sigma_1^x\sigma_2^x + \sigma_1^y\sigma_2^y)
+
g_z\sigma_1^z\sigma_2^z
+
B_0\,\sigma_1^z,
\ee
where $\vec{\sigma}_i\equiv(\sigma_i^x,\sigma_i^y,\sigma_i^z)^T$ are
Pauli matrices for the $i-$th spin, $\vec{B}$ is the external magnetic
field acting simultaneously on both spins, anisotropically
(isotropically) if $\gamma_1 \neq \gamma_2$ ($\gamma_1 = \gamma_2$),
$g$ describes the $xy$ coupling, $B_0$ is an offset magnetic field
applied only to the first spin, breaking the exchange symmetry if
non-zero, and $g_z$ indicates the interaction in the $z$ direction,
which can turn the system into the $SU(2)$ Heisenberg Hamiltonian for
the choice of constants $g_z=1$, $g=2$,
$\gamma_{1}=\gamma_{2}=1$ and $B_{0}=0$.

In the present analysis, we will fix $\gamma_1$, $\gamma_2$, $g$,
$B_0$ and $g_z$ and restrict ourselves to consider the Berry curvature
with respect to the external applied magnetic field $\vec{B}\in
\mathbb{R}^3$, defining our parameter space. The vector $\vec{B}$ will
interchangeably be written in spherical $(B,\theta,\phi)$ or Cartesian
$(B_x, B_y, B_z)\equiv (x,y,z)$ coordinates, whichever is more
convenient. The term $g$ merely sets the energy scale, and so we will
consider units in which $g=2$ from here onward.
%
% figures should be put into the text as floats.
% Use the graphics or graphicx packages (distributed with LaTeX2e)
% and the \includegraphics macro defined in those packages.
% See the LaTeX Graphics Companion by Michel Goosens, Sebastian Rahtz,
% and Frank Mittelbach for instance.
%
% Here is an example of the general form of a figure:
% Fill in the caption in the braces of the \caption{} command. Put the label
% that you will use with \ref{} command in the braces of the \label{} command.
% Use the figure* environment if the figure should span across the
% entire page. There is no need to do explicit centering.
%

The eigenenergies of (\ref{eq:orighamiltonian}) possess azimuthal
symmetry, since the Hamiltonian and ground-state at arbitrary $\phi$
are trivially connected to their expressions at $\phi=0$.
In other words,
$H(B,\theta,\phi) = R^\dag(\phi) H(B,\theta,0) R(\phi)$, where
$R(\phi) = \exp(i\, \phi\, \sigma^z_{\mathrm{tot}}/2)$, and similarly
for the ground-state,
$\ket{\Psi_0(B,\theta,\phi)} = R^\dag(\phi) \ket{\Psi_0(B,\theta,0)}$.
The Hamiltonian is real at $\phi=0$, and therefore a gauge choice is
made requiring the eigenfunctions to be real-valued.
As a consequence of this gauge, the components $A_B$ and $A_\theta$ of
the Berry connection vanish, and the only non-zero
component $A_\phi$ can be calculated explicitly (see Appendix \ref{app:A})\label{!!! - SUP A}
\be
\label{eq:bconnection}
\vec{A}
=
\frac{1}{B \sin\theta}
\frac{\left\langle\sigma^z_{\mathrm{tot}}\right\rangle}{2}
\hat{\phi}\,. 
\ee
One can thus use this result to experimentally measure the Berry
connection by measuring the ground-state expectation value of the
total magnetization, with the Berry curvature obtained by taking the
curl of Eq.~(\ref{eq:bconnection}).

In analogy with E\&M, one of Maxwell's equations in $\mathbb{R}^3$ for
the vector field $\vec{F}$ is
\begin{align}
\label{eq:divf}
& \vec{\nabla} \cdot \vec{F} = 2\pi \rho_{m},
\end{align}
with $\rho_{m}$ denoting the effective magnetic charge density.
The expression above is nothing but the differential form of
Eq.~(\ref{eq:chnum}), showing that the divergence of $\vec{F}$ is
equal to the effective magnetic charge (first Chern number). The role
of Chern numbers as topological quantifiers in quantum systems has
been widely investigated, and it is still a very active
field~\cite{Wen2004,Bernevig2013}. A direct measurement of the Berry
curvature was proposed in~\cite{Gritsev2012, Avron2011}, where it was
shown to be given by the non-adiabatic response of certain physical
observables. This was experimentally confirmed with systems of
superconducting qubits~\cite{Schroer2014, Roushan2014}, where the
first Chern number quantization was readily confirmed.

However, the role of $\vec{\nabla}\times\vec{F}$ has not been explored
so far. In three-dimensional space, any vector field is uniquely
represented by its divergence and curl. The divergence of $\vec{F}$,
as seen from Eq.~(\ref{eq:divf}), is given by effective magnetic
charges, while the curl is analogous to ``electric" currents.
In what follows, we then study in detail the divergence and curl of
$\vec{F}$ for different fixed set of values of the parameters
$\gamma_1,\gamma_2,g_z$ and $B_0$. We start with the choice that makes
the Hamiltonian~(\ref{eq:orighamiltonian}) $SU(2)$ symmetric, and
break symmetries in each subsequent case.

%%%%%%%%%%

\section{Heisenberg interaction}
\label{Section: Heisenberg interaction}
%%%%%%%%%%%

The simplest system extending the
aforementioned E\&M analogy to continuous magnetic charge densities
has the Hamiltonian
$H
=
\vec{B} \cdot (\vec{\sigma}_1 + \vec{\sigma}_2) + \vec{\sigma}_1\cdot
\vec{\sigma}_2$.
A similar system and its charge configuration was studied
in~\cite{Oh2009}.
Ours corresponds to the two-spin Heisenberg model in an external
$\vec{B}$ field, possessing $SU(2)$ symmetry.
It is obtained from the Hamiltonian (\ref{eq:orighamiltonian}) by
setting the parameters to $\gamma_1=\gamma_2=g_z=1$, $B_0=0$.
\begin{figure}[]
	\includegraphics[scale=0.49,trim=1.5mm 0mm 3mm 1mm, clip]{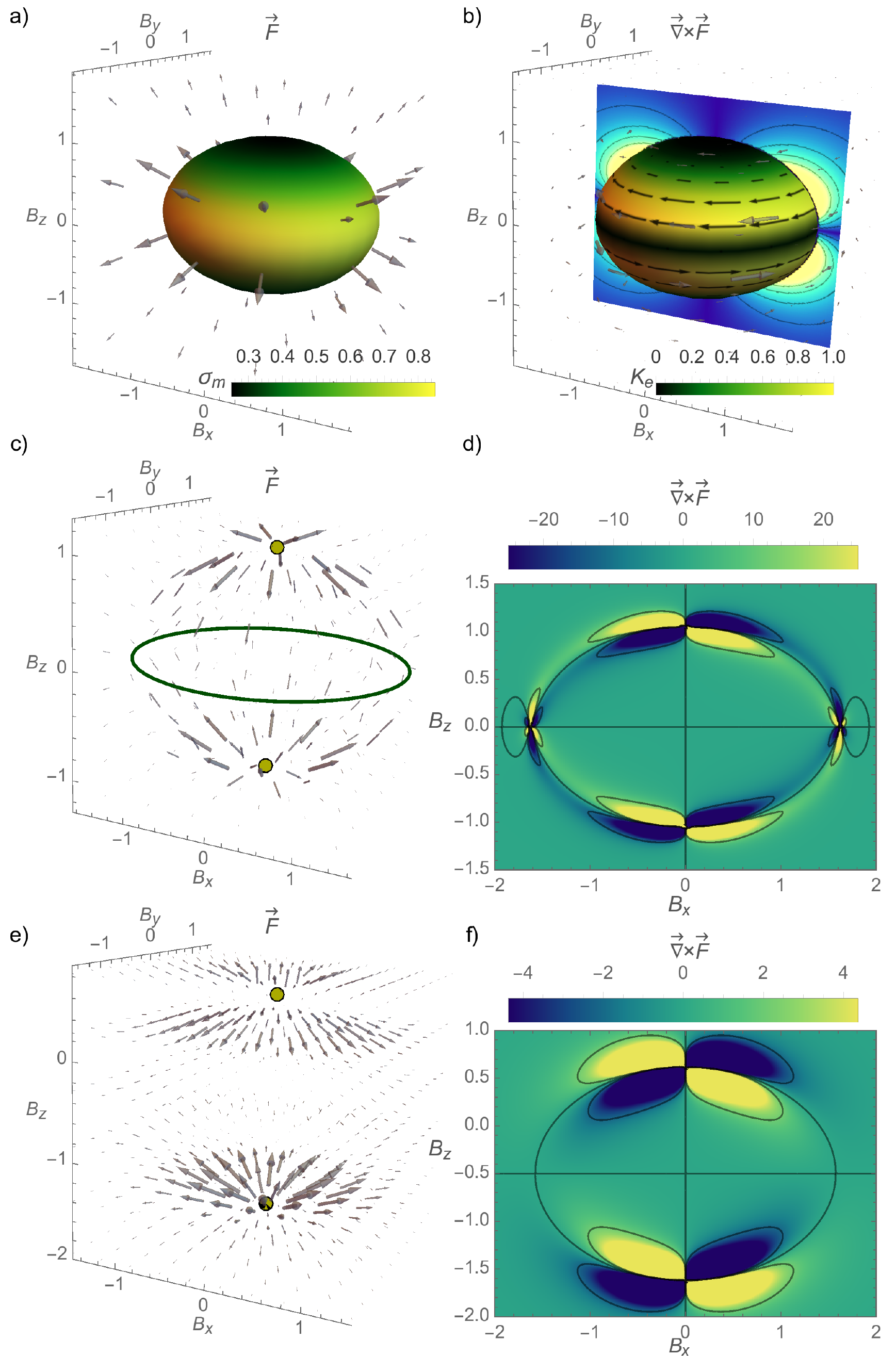}
	\caption{\textit{XXZ-interaction} ($g_z=0$):
		 a) Berry curvature $\vec{F}$ (light arrows) and
                 magnetic surface charge density  $\sigma_m$  (color
                 bar).
		b) $\vec{\nabla}\times\vec{F}$, with magnitude shown
                in the $xz-$plane. The curl has only a $\phi$
                component, and the colors on the ellipsoid illustrate
                the magnitude of the electric surface current density
                $\vec{K}_e=K_e \hat{\phi}$, with direction indicated
                by the darker arrows.
		\textit{Anisotropic fields} ($g_z=0$, $\alpha = 0.3$):
		c) Berry curvature $\vec{F}$ (light arrows), showing
                two charges (yellow dots) on the $z-$axis, plus an
                uncharged ring (green) in the $xy-$plane.
		d) The curl of the Berry curvature, shown only in the
                $xz-$plane since it
                has azimuthal symmetry.
		\textit{Broken exchange symmetry} ($g_z=0$, $\alpha =
                0$, $B_0 = 1$):
		e) Berry curvature $\vec{F}$ (light arrows) showing
                now the only two magnetic charges on the
                $z-$axis.
		f) $\vec{\nabla} \times \vec{F}$ shown in the $xz-$plane.
		\label{fig:degeneraciescurl}
		}
\end{figure}

The ground-state degenerates on the sphere of radius $B=2$, dividing
the parameter space into two disjoint regions.
The Berry curvature in this case is
\be
\label{eq:heisenbcurv}
\vec{F}
=
\begin{cases}
 0, \quad B<2, \\
\frac{1}{2}q_m\frac{\hat{B}}{B^2}\,, \quad B>2\, ,
\end{cases}
\ee
where $q_m=2$ gives the effective magnetic charge (see Appendix \ref{app:B})\label{!!! - SUP B}.
The effective magnetic field defined by the Berry curvature above is
akin to the electric field of a hollow conducting sphere of radius
two. The total magnetic charge is equal to the Chern number,
$\text{ch}_1 = 2$, and can be obtained from Eq.~(\ref{eq:chnum}). The
magnetic charge density distribution $\rho_m$ is uniform since the
sphere is a surface of constant curvature. The curl of $\vec{F}$ is
equal to zero since the field falls of radially as $1/B^2$. This will
not be the case in the following examples.

\section{XXZ interaction}
\label{Section:XXZ interaction}
%%%%%%%%%%%%%%
%%%%%%%%%%%%%%
Let us now consider the case where $g_z
\neq 1$, and as before, $\gamma_1=\gamma_2=1$, $B_0 = 0$.
Unlike the Heisenberg case, 
if $|g_{z}|<1$
($|g_z|>1$) we find that the $SU(2)$ symmetry is broken, and the
charged sphere of the prior case gets continuously squeezed
(stretched) along the $z-$axis, becoming an oblate (prolate) ellipsoid
of revolution. In analogy to the charge distribution on conductors in
electrostatics, the magnetic charge density is no longer uniformly
distributed, but accumulates in regions of higher curvature (see
Fig.~\ref{fig:degeneraciescurl}a). In spite of the non-uniform surface
charge density, the total charge on the entire surface remains the
same as for the previous case ($\text{ch}_1=2$). This can be concluded
from the fact that the ground-state remains fully polarized at large
$B$, yielding the total effective charge enclosed as a topologically
protected integer equal to $\text{ch}_1 = 2$.

Figure~\ref{fig:degeneraciescurl}b shows the existence of a surface
current $\vec{K}_e \neq 0$ defined by the discontinuity of the
parallel component of $\vec{F}$ across the surface, which implies that
$\vec{\nabla} \times \vec{F} \neq 0$ (see Appendix \ref{app:C})\label{!!! - SUP C}. The Berry
curvature has only $\hat{B}$ and $\hat{\theta}$ components, and
therefore its curl is parallel to $\hat{\phi}$. The non-uniform
magnetic charge distribution produces a quadrupole in the curl of
$\vec{F}$.

In the previous two cases we have explored situations of high
symmetry, where the magnetic  charges occur as surface densities
spread on closed degeneracy surfaces, instead of the more commonly
studied discrete monopole charges~\cite{Berry1984}. Similar cases of
continuous surfaces with magnetic charge densities have been explored
elsewhere~\cite{Oh2009}. The newest aspect of the aforementioned
results is shown by the curl of the Berry curvature, which displays a
characteristic quadrupole pattern.

\section{Bypassing degeneracy crossings}
\label{Section: Bypassing degeneracy crossings}
The points belonging to the closed surfaces in the two previous cases indicate the locus in parameter space where there are degeneracies in the ground-state.
Interestingly, inside all the previous surfaces, the ground-state is a singlet $\ket{\Psi_0} \equiv \frac{1}{\sqrt{2}}(\left\rvert\uparrow\downarrow\right\rangle - \left\rvert\downarrow\uparrow\right\rangle)$, i.e., a Bell entangled state of the two qubits (see Appendixes \ref{app:B} \& \ref{app:C})\label{!!! - SUP B-C}.
Equation~(\ref{eq:bconnection}) then implies a vanishing Berry
connection and curvature; in the region outside the closed surfaces,
$\ket{\Psi_0}$ has contributions of the other vectors in the
spin-product basis. At first sight, it might seem impossible to
experimentally start with a high polarizing field $B_z \gg B_x \approx
0$ where $\ket{\Psi_{0}} \approx
\left\rvert\uparrow\uparrow\right\rangle$ to subsequently prepare
adiabatically a pure singlet-state without crossing the continuous
degeneracy surface, which would introduce excitations and break the
adiabaticity. 
\begin{figure*}[]
	\includegraphics[scale=0.53,trim=0.05mm 2.5mm 3.5mm 1.5mm, clip]{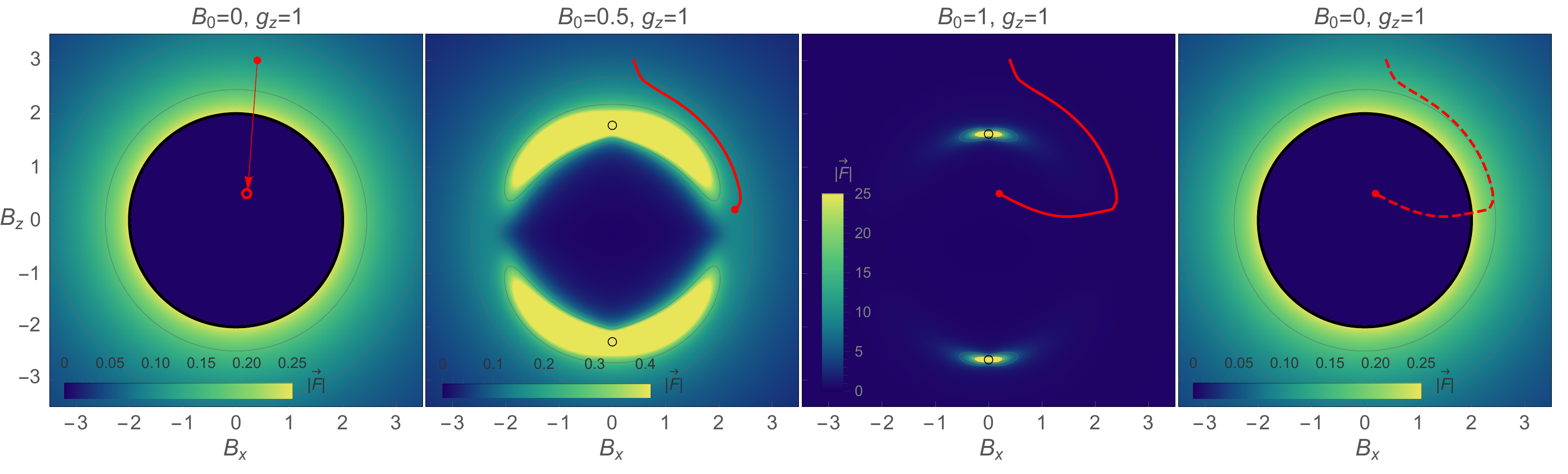}
		\caption{\textit{Opening an adiabatic passage between
                    topologically disjoint regions in parameter
                    space.} First panel: starting with a point in
                  parameter space corresponding to a state outside the
                  sphere defined by the Heisenberg Hamiltonian. Second
                  and third panel: Breaking symmetry by adiabatically
                  introducing a pinning field in one of the spins
                  creates a different topological magnetic charge
                  distribution (with the total magnetic charge
                  conserved). One can now evolve the system to a
                  previously adiabatically inaccessible region. Fourth
                  panel: Reintroducing the symmetry by removing the
                  symmetry-breaking term allows one to bypass the
                  continuous crossing surface and enter a previously
                  adiabatically disconnected region in parameter space.
		\label{fig:adiabaticpassage}
		}
\end{figure*}

In order to bypass this topological constraint, we now consider
situations with significantly reduced symmetry, and we observe a sharp
collapse of the surface charge density to the more familiar case of
magnetic monopoles.
This singular change in the topology of the monopole charge density is
unlike anything in classical E\&M, and we explore this phase
transition to open a passage and access the interior regions of the
previous cases by adiabatically breaking and reestablishing symmetries
(see Fig.~\ref{fig:adiabaticpassage}).
We also show how the effective electromagnetic fields respond to this
transition.
\section{Anisotropic fields}
\label{Section: Anisotropic fields}
%%%%%%%%%%%%%%%%
%%%%%%%%%%%%%%%%
Anisotropy is introduced by setting
$\gamma_1 = 1+\alpha$, $\gamma_2 = 1-\alpha$ with $\alpha \neq 0$ and
$g_z = B_0 = 0$. The introduction of $\alpha$ breaks the symmetry that
allows the existence of a two-dimensional manifold where the
ground-state degeneracies occurs. The previous surfaces now collapse
to two points, located on the $z-$axis at $\pm
g/(2\sqrt{(1-\alpha^2)}$ due to azimuthal symmetry of the eigenenergies.
These two points correspond to energy level crossings in the
ground-state manifold and act like sources of $\vec{F}$. The total
Chern number in the entire parameter space is topologically protected
and equal to $+2$, therefore each source carries an effective magnetic
charge equal to $+1$.

These magnetic monopoles are visible as singularities in the
divergence of the Berry curvature, which is unsurprisingly zero away
from these two singularities.
For the curl of the Berry curvature, we find a quadrupolar field
pattern very similar to what we saw in the previous section for the
charged surface.
However, surprisingly we also find the appearance of two other points
on the $x$-axis.
Respecting the model's symmetry by revolving the plotted planes around
the $z-$axis, one can see that these points in fact correspond to a
ring of degeneracies, centered at the origin in the $xy-$plane, with
radius $\varrho = \sqrt{2(1+\alpha^2)}/(1-\alpha^2)$ (see
Fig.~\ref{fig:degeneraciescurl}c).
Most interestingly, this ring is uncharged as can be inferred from a
topological argument: the total Chern number for the entire parameter
manifold must remain equal to $+2$, and the monopoles on the $z-$axis
each carries a unit charge, as can be verified using Gauss law.
The Berry curvature in its vicinity exhibits a saddle-point behavior,
rather than acting like a sink or source for the $\vec{F}$ vector
field. The analogous configuration in E\&M are two electric charges
with a conducting ring placed halfway in between, which introduces
boundary conditions for the fields. As can be seen in
Fig.~\ref{fig:degeneraciescurl}c, the curl of Berry curvature shows a
hexapole pattern for the intersections on the $xy-$plane. Thus, the
presence of the uncharged ring, although not obvious from the Berry
curvature field, can clearly be observed in the
$\vec{\nabla}\times\vec{F}$ graph, as they exhibit a distinct pattern
compared to degeneracies having an effective charge. The curl then
apparently contains additional geometric information about the
ground-state manifold of the system, which has not been explored so
far.

It is readily confirmed that the gap vanishes at this ring of
singularities despite the absence of effective magnetic charge.
Interestingly, crossing this degeneracy by fixing $B_x=\varrho$ and varying
$B_z$, we find that the energies exhibit a quadratic touching, which
in the chemistry literature is known as Renner-Teller intersection
points~\cite{Zwanziger:1987,Yarkony:1996wx,Yarkony:1998ge,Oh2009},
fundamentally different from conical intersections since they do not
give rise to a geometric phase, and consequently have a Chern number
equal to $0$.
This quadratic touching comes from a symmetry of the Hamiltonian,
namely $B_z \to -B_z$, and therefore is not present for curves that do
not cross the degeneracy vertically, e.g., the energy levels for fixed
$B_z=0$ when varying $B_x$.

\section{Broken exchange symmetry}
\label{Section: Broken exchange symmetry}
For the final case, consider
$\gamma_1 = \gamma_2 =1$, $g_z = 0$ and $B_0 =1$.
Due to the pinning field $B_0$ on the first spin the exchange symmetry
between the two spins
is broken, and only the azimuthal symmetry in the eigenenergies is
left.
The crossing points now lie solely on the $z-$axis, with the two
monopoles located at $B^{(\pm)}_{z}\equiv(-B_0 \pm \delta)/2$, where
$\delta \equiv \sqrt{B_0^2 + g^2}$ is the degenerate ground-state
energy for the case in consideration.
The curl of $\vec{F}$ for this case shows a persistent quadrupole
pattern around the crossing points in the $z-$axis, although one
observes a bending of the lobes toward each other, which increases
with $B_0$ (see Fig.~\ref{fig:degeneraciescurl}e). The point charges
along the $z-$axis are no longer symmetric with respect to the
$xy-$plane, and their location varies as a function of $B_0$, given by
$B_z^{(\pm)}$.

\section{Curl analysis}
\label{Section: Curl analysis}
To understand the curl behavior of $\vec{F}$ analytically, we
calculate the Berry connection for the broken exchange symmetry case
at the degeneracy points $B_{z}^{(\pm)}$ using perturbation theory and
obtain $\vec{\nabla}\times\vec{F}$ around $B_{z}^{(\pm)}$ (see Appendix \ref{app:E})\label{!!! - SUP E}.
The leading order expression for the curl is given by
\be
\label{eq:pertfcurl}
\vec{\nabla} \times \vec{F}_{(\pm)}
=
-
\frac{3}{4}
\frac{\beta^2 \sin 2\vartheta}{\gamma^2(1-\beta^2\sin^2\vartheta)^{5/2}}
\frac{1}{dB^3} \, \hat{\phi} + \cdots \, ,
\ee
where $\beta^2 \equiv g/\delta$, $\gamma \equiv 1/\sqrt{(1-\beta^2)}$, 
and with respect to the coordinate system centered at the monopole
$B_{z}^{(\pm)}$ respectively, given in spherical coordinates
$(dB, \vartheta, \phi)$.
This expression reproduces qualitatively the quadrupole seen
numerically (see Fig.~\ref{fig:degeneraciescurl}d and Appendix \ref{app:F}\label{!!! - SUP F}) 
and suggests the possibility that this pattern may be robust in the
vicinity of Berry curvature sources.

%%%%%%%%%%%%%
%%%%%%%%%%%%%%%%%%%%%%%%%%%%%%%%%%%%%%%%%%%%%%%%%%%%%%%%%%%%%%%%%%%%%%%%%%
%%%%%%%%%%%%%%%%%%%%

\label{Section: Conclusion}
\section{Conclusion}
The analogy between E\&M and degeneracies in quantum systems has been outlined many years ago, and it is still a field of active research, mainly due to its applications to adiabatic quantum computing and the recent burst of interest in topological transitions. 
For highly symmetric systems, we have shown how symmetry-breaking perturbations allows one to open adiabatic passages in previously topologically disjoint regions, thereby allowing the full parameter space to be explored. 
The procedure outlined in this paper is general and robust, and not necessarily restricted to qubits. We note that by identifying angles of the magnetic field with quasi-momenta, the two-spin system here analyzed can be mapped to a four-band model of a topological insulator with a rich phase diagram, similar to the construction in~\cite{Roushan2014}. Therefore results presented in this paper can find direct analogues in other systems.
The system in analysis was chosen as a good illustrative example due to experimental feasibility of measuring the Berry connection $\vec{A}$ by relating it to the ground-state expectation value of the total magnetization. 
The Berry curvature and its curl can then be experimentally obtained, and the results here presented can be verified.

We also highlighted the existence of degeneracy points with vanishing Chern number, and exemplified how they fit within the E\&M analogy as boundary conditions for the $\vec{F}$ field. 
Finally, the curl of the Berry curvature was explored, with different behavior near charged and uncharged points, indicating the possibility that this quantity might carry geometrical information about the ground-state manifold previously unexplored.

% Specify following sections are appendices. Use \appendix* if there
% only one appendix.

% If you have acknowledgments, this puts in the proper section head.
\begin{acknowledgments}
\label{Section: Acknowledgments}
The authors thank P. Roushan, A. Dunsworth and T. Pudlik for enlightening discussions. T. S. and A. P were supported by AFOSR FA9550-13-1-0039, NSF DMR-1506340, ARO W911NF1410540. M. T. was funded by the Swiss National Science Foundation (SNSF), M. K. was supported by Laboratory Directed Research and Development (LDRD) funding from Berkeley Lab, provided by the Director, Office of Science, of the U.S. Department of Energy under Contract No. DE-AC02-05CH11231.
\end{acknowledgments}
\label{APPENDIX BEGINS HERE!}
\begin{appendix}
\section{Locations of the effective magnetic charges for an interacting two qubit system}
\label{app:A}
In the main text, we plot the Berry curvature and its curl 
for the ground-state $\ket{\Psi_{0}}$ to illustrate the locations of the
ground-state degeneracies in parameter space.
We use the analogy with electromagnetism,
pointed out by M.\ V.\ Berry~\cite{Berry1984},
that identifies the Berry curvature $\vec{F}$ with an effective
magnetic field in parameter space whose vector potential is the Berry
connection
$\vec{A} = i \bra{\Psi_{0}} \vec{\nabla} \ket{\Psi_{0}}$.
The locations of the associated magnetic charges are
given by the ground-state degeneracies, and their charge is determined
by the first Chern number.
The fact that degeneracies of the ground-state act as magnetic
charges can be seen by the following reasoning:
the vector identity $\vec{\nabla}\cdot(\vec{\nabla}\times\vec{A})=0$
holds only if $\vec{A}$ has continuous derivatives.
This is no longer the case when the ground-state becomes degenerate,
since at these points $\ket{\Psi_{0}}$ undergoes
a discontinuous change and so the derivatives of $\vec{A}$ become
discontinuous.
As a result, at the degeneracies we have
$\vec{\nabla}\cdot(\vec{\nabla}\times\vec{A}) \neq 0$,
and in analogy with Maxwell's equations we can write an equivalent
Gauss's law for the Berry curvature
\be
\label{eq:gauss}
\vec{\nabla}\cdot\vec{F}=2\pi\rho_{m},
\ee
where $\rho_{m}$ is the effective magnetic charge density.
The volume integral of~(\ref{eq:gauss}) yields
\be
\label{eq:intgauss}
\iint_{\Sigma} \vec{F} \cdot d\vec{S} = 2\pi \iiint_{V} \rho_{m} dV,
\ee
where the divergence theorem was applied to the left hand side of the
equation.
According to the Chern theorem~\cite{Nakahara2003}, the integral of the Berry
curvature over a closed manifold $\Sigma$ is quantized in units of
$2\pi$, and this number defines the first Chern number
\be
\mathrm{ch}_{1} = \frac{1}{2\pi} \iint_{\Sigma} \vec{F} \cdot
d\vec{S}.
\ee
The comparison of the previous two equations implies the quantization 
of $\iiint_{V} \rho_{m} dV$, which also defines
the effective charge enclosed by the manifold $\Sigma$.

For a single magnetic monopole charge $q_{m}$, the magnetic charge
density is $\rho_{m}= q_{m} \, \delta(\vec{r})$ and the associated magnetic
field is then given, in view of Eq.~(\ref{eq:intgauss}), by
\be
\label{eq:berrycsohalf}
\vec{F}=\frac{1}{2} q_{m} \frac{\hat{r}}{|\vec{r}|^{2}}, 
\ee
where the prefactor of $1/2$ sets the units such that the charge $q_{m}$
is equal to the Chern number.
This example is realized by a single qubit (spin-$1/2$) in an external
magnetic field $\vec{B}$, where the resulting Berry curvature is given
by~(\ref{eq:berrycsohalf}) and therefore analogous to an effective
magnetic field in parameter space $(B_{x},B_{y},B_{z})$, or $\vec{r} \equiv \vec{B}$, created by a
magnetic monopole sitting at $B=0$ and carrying a charge $q_{m}=1$.

Finally, we note that the Berry curvature $\vec{F}$ associated with
the ground-state can also be rewritten, using the resolution of the
identity $\sum_{m}\ket{\Psi_{m}}\bra{\Psi_{m}}=1$, 
as a sum over all other eigenstates
\be
\label{eq:berrycexpsum}
\vec{F}
=
i
\sum_{m \neq 0}
\frac{\bra{\Psi_{0}} \vec{\nabla} H \ket{\Psi_{m}} 
      \times
      \bra{\Psi_{m}} \vec{\nabla} H \ket{\Psi_{0}}}
{(E_{0}-E_{m})^{2}}.
\ee
This equation highlights that degeneracies in the
ground-state, $E_{0}=E_{m}$, act as charges for $\vec{F}$.
In particular, the expression~(\ref{eq:berrycexpsum}) is useful to
compute the Berry curvature numerically, if the Hamiltonian is not
analytically diagonalizable.

In the following, we illustrate the calculations that lead to the
localization of the ground-state degeneracies in parameter space for
the two-qubit systems considered in the main text.
We also calculate the corresponding Berry connection,
curvature and its curl.
First, however, let us review some important properties of the
system studied in the main text, consisting of two interacting qubits,
with each qubit separately coupled to external magnetic fields.
The Hamiltonian of this system is given by
\beq
\label{eq:sysham}
H
=
\vec{B} \cdot \lp \gamma_{1} \vec{\sigma}_{1} + \gamma_{2} \vec{\sigma}_{2} \rp
+
\frac{g}{2} \lp \sigma_{1}^{x} \sigma_{2}^{x} + \sigma_{1}^{y}\sigma_{2}^{y} \rp
+
\nn \\
+g_{z} \sigma_{1}^{z}\sigma_{2}^{z}
+
B_{0}\, \sigma_{1}^{z},
\eeq
where $\vec{\sigma}_i\equiv(\sigma_i^x,\sigma_i^y,\sigma_i^z)^T$
are the usual Pauli matrices for the $i-$th spin
\be
\sigma_{i}^{x}
=
\begin{pmatrix}
 0 & 1 \\
 1 & 0
\end{pmatrix},
~~
\sigma_{i}^{y}
=
\begin{pmatrix}
 0 & -i \\
 i & 0
\end{pmatrix},
~~
\sigma_{i}^{z}
=
\begin{pmatrix}
 1 & 0 \\
 0 & -1
\end{pmatrix},
\ee
with $i=1,2$.
The external magnetic field is
$\vec{B} = (B_{x},B_{y},B_{z})^{T} \equiv (x,y,z)^{T}$,
which acts isotropically on both spins if $\gamma_{1}=\gamma_{2}$, and
anisotropically if $\gamma_{1}\neq\gamma_{2}$.
The field $B_{0}$ is a local magnetic 
field applied only to the first spin in the $z$ direction,
and allows us to break the exchange symmetry between the two
spins.
The term $g$ is the energy scale of the interaction between the two spins in
the $x$ and $y$ direction, and $g_{z}$ indicates the interaction in the
$z$ direction.

As mentioned in the main text, we consider the parameters $\gamma_1$,
$\gamma_2$, $g$, $B_0$ and $g_z$ as fixed and restrict
ourselves to the case of an adiabatically varying external magnetic field
$\vec{B}$ that spans the parameter space
$\mathcal{M} \equiv \mathbb{R}^{3}$.
The magnetic field $\vec{B}$ in spherical coordinates
$(B,\theta,\phi)$ reads 
$\vec{B}
=
B \, (\sin\theta \cos\phi, \sin\theta \sin\phi, \cos\theta)^{T}
=
B \, \hat{B}(\theta,\phi)$,
where
$\hat{B}(\theta,\phi)$ is the unit vector in the radial direction.
The Hamiltonian in spherical coordinates can be rewritten as
\beq
H(B, \theta, \phi)
=
B \, \hat{B}(\theta, \phi) \cdot \lp \gamma_{1} \vec{\sigma}_{1} + \gamma_{2} \vec{\sigma}_{2} \rp
+
\qquad
\nn \\
+
\frac{g}{2} \lp \sigma_{1}^{x} \sigma_{2}^{x} + \sigma_{1}^{y}\sigma_{2}^{y} \rp
+
g_{z} \sigma_{1}^{z}\sigma_{2}^{z}
+
B_{0} \, \sigma_{1}^{z},
\eeq
and written in this form, it is evident that the Hamiltonian at
arbitrary $\phi$ can be obtained from the one at $\phi=0$ by
\be
\label{eq:hamiprop}
H(B, \theta, \phi)
=
R^\dag(\phi) H(B,\theta,0) R(\phi),
\ee
where $R(\phi)=\exp(i\, \phi\, \sigma_{\mathrm{tot}}^{z}/2)$ and
$\sigma_{\mathrm{tot}}^{z} = \sigma_{1}^{z}+\sigma_{2}^{z}$.
Equation~(\ref{eq:hamiprop}) implies that the eigenstates of
$H(B, \theta, \phi)$ are simply given by a rotation of the eigenstates
of $H(B, \theta, 0)$,
\be
\ket{\Psi_{m}(B,\theta,\phi)}
=
R^{\dag}(\phi) \ket{\Psi_{m}(B,\theta,0)},
\ee
and that the eigenenergies of $H(B, \theta, \phi)$ and $H(B, \theta, 0)$
are the same, $E_{m}(B,\theta,\phi)=E_{m}(B,\theta,0)$.
Note that Eq.~(\ref{eq:hamiprop}) does not provide any additional
conservation laws but it is useful for the calculation of the Berry
connection and curvature.

We emphasize that the Berry connection is a connection one-form on the
parameter space $\mathcal{M}$, in general defined by
$A \equiv i \bra{\Psi_{0}}\mathrm{d}\ket{\Psi_{0}}$,
where $\mathrm{d}$ is the exterior derivative.
The corresponding Berry curvature is a two-form defined by
$F=\mathrm{d}A$.
In local coordinates $(x^{1}, x^{2}, \ldots)$ we can write
\be
A = A_{\mu}dx^{\mu},
~
A_{\mu} = i \bra{\Psi_{0}}\pd_{\mu}\ket{\Psi_{0}},
\ee
and where $\pd_{\mu} = \frac{\pd}{\pd x^{\mu}}$, $\mu=1,2,\ldots$.
Similarly, the Berry curvature in local coordinates reads
\be
F = \frac{1}{2} F_{\mu\nu} dx^{\mu} \wedge dx^{\nu},
~
F_{\mu\nu} = \pd_{\mu} A_{\nu} - \pd_{\nu} A_{\mu},
\ee
where $dx^{\mu} \wedge dx^{\nu}$ is the wedge product of two one-forms.
For a three dimensional parameter space $\mathcal{M}$ as the one 
studied in the main text, the components of the Berry connection
one-form $A_{\mu}$ can be collected in a vector
as $\vec{A} = i \bra{\Psi_{0}}\vec{\nabla}\ket{\Psi_{0}}$.
Similar, the Berry curvature two-form can be mapped to a vector
through $\vec{F}=\vec{\nabla}\times\vec{A}$.
This mapping can be seen explicitly from the definition of the Berry curvature
in local coordinates $F_{\mu\nu} = \pd_{\mu} A_{\nu} - \pd_{\nu} A_{\mu}$, which 
is an antisymmetric tensor. In three dimensions, it reduces to
\be
(F_{\mu\nu})
=
\begin{pmatrix}
F_{11} & F_{12} & F_{13} \\
F_{21} & F_{22} & F_{23} \\
F_{31} & F_{32} & F_{33} 
\end{pmatrix}
\equiv
\begin{pmatrix}
0 & F_{3} & -F_{2} \\
-F_{3} & 0 & F_{1} \\
F_{2} & -F_{1} & 0
\end{pmatrix}
\ee
and thus we can write $\vec{F}=(F_{1},F_{2},F_{3})^{T}=(F_{23},F_{31},F_{12})^{T}$.

In view of the discussion in the previous paragraph the Berry
connection in Cartesian coordinates, $(B_{x},B_{y},B_{z}) \equiv
(x,y,z)$, reads
\be
\vec{A}^{(C)}(x,y,z)
=
A_{x}\hat{x} + A_{y}\hat{y} + A_{z}\hat{z},
\ee
where $A_{\mu} = i \bra{\Psi_{0}}\pd_{\mu}\ket{\Psi_{0}}$, with
$\mu=\{x,y,z\}$, $\pd_{\mu}=\frac{\pd}{\pd \mu}$, and $\hat{x}$, $\hat{y}$
and $\hat{z}$ are the unit vectors in Cartesian coordinates. 
One should be wary, since there is a potential for ambiguity in this
notation if a different choice of coordinate system is considered.
For example, with respect to spherical coordinates $(B,\theta,\phi)$, the Berry
connection becomes
\be
\vec{A}^{(S)}(B,\theta,\phi)
=
A_{B}\hat{B} + A_{\theta}\hat{\theta} + A_{\phi}\hat{\phi},
\ee
where now we must define 
\beq
A_{B} &=& i \,\ \bra{\Psi_{0}}\pd_{B}\ket{\Psi_{0}}
\nn
\\
A_{\theta} &=& i \, \frac{1}{B} \, \bra{\Psi_{0}}\pd_{\theta}\ket{\Psi_{0}}
\nn
\\
A_{\phi} &=& i \, \frac{1}{B \sin\theta} \, \bra{\Psi_{0}}\pd_{\phi}\ket{\Psi_{0}}
\eeq
since in spherical coordinates the operator $\vec{\nabla}$
is given by 
\be
\vec{\nabla} f
=
\frac{ \partial f}{ \partial B} \hat{B}
+
\frac{1}{B}\frac{ \partial f}{ \partial \theta}\hat{\theta}
+
\frac{1}{B \sin\theta}\frac{\partial f}{\partial \phi}\hat{\phi},
\ee
where 
\beq
\hat{B}
&=&
\sin \theta \cos \phi \, \hat{x}
+
\sin \theta \sin \phi \, \hat{y}
+
\cos \theta \, \hat{z}
\nn
\\
\hat{\theta}
&=&
\cos \theta \cos \phi \, \hat{x}
+
\cos \theta \sin \phi \,\hat{y}
-
\sin \theta \, \hat{z}
\nn
\\
\hat{\phi}
&=&
-\sin \phi \, \hat{x} + \cos \phi \, \hat{y}
\eeq
are the local orthogonal unit vectors in the directions of increasing 
$B$, $\theta$ and $\phi$, respectively.
Note that the Cartesian unit vectors can be expressed as
\beq
\hat{x}
&=&
\sin\theta \cos\phi \, \hat{B}
+
\cos\theta \cos\phi \, \hat{\theta}
-
\sin\phi \, \hat{\phi}
\nn
\\
\hat{y}
&=&
\sin\theta \sin\phi \, \hat{B}
+
\cos\theta \sin\phi \, \hat{\theta}
+
\cos\phi \, \hat{\phi}
\nn
\\
\hat{z}
&=&
\cos\theta \, \hat{B}
-
\sin\theta \, \hat{\theta}
\eeq
or the spherical unit vectors as
\beq
\hat{B}
&=&
\frac{x\,\hat{x} + y\,\hat{y} + z\,\hat{z}}{\sqrt{x^{2}+y^{2}+z^{2}}}
\nn
\\
\hat{\theta}
&=&
\frac{x z\,\hat{x} + y z\,\hat{y} - (x^{2}+y^{2})\,\hat{z}}{\sqrt{x^{2}+y^{2}}\sqrt{x^{2}+y^{2}+z^{2}}}
\nn
\\
\hat{\phi}
&=&
\frac{-y\,\hat{x} + x\,\hat{y}}{\sqrt{x^{2}+y^{2}}}.
\eeq
The Berry phase, given by the integral of the Berry connection along a
closed loop $\mathcal{C}$ in parameter space, can be written as
\be
\gamma
=
\int_{\mathcal{C}} \vec{A}^{(C)} \cdot d\vec{B}
=
\int_{\mathcal{C}} \vec{A}^{(S)} \cdot d\vec{B},
\ee
where $d\vec{B}=dx\, \hat{x}+dy\, \hat{y}+dz\, \hat{z}$ in Cartesian
coordinates and
$d\vec{B} = dB\, \hat{B} + B \, d\theta \, \hat{\theta} + B \sin\theta\, d\phi\, \hat{\phi}$ 
in spherical coordinates.

The relation
$\ket{\Psi_{0}(B,\theta,\phi)}=R^{\dag}(\phi) \ket{\Psi_{0}(B,\theta,0)}$,
where
$R(\phi)=\exp(i\, \phi\, \sigma_{\mathrm{tot}}^{z}/2)$,
allows us to calculate the Berry connection in spherical coordinates
straightforwardly.
First, observe that the quantities
$\bra{\Psi_{0}}\pd_{\mu}\ket{\Psi_{0}}$, for $\mu=\{ B,\theta,\phi \}$, must
be purely imaginary numbers.
This can be seen by differentiating the
normalization condition $\bracket{\Psi_{0}}{\Psi_{0}}=1$ with respect
to either $B,\theta$ or $\phi$.
A gauge choice allows us to choose the eigenstates of the
Hamiltonian~(\ref{eq:sysham}) to be real at $\phi=0$,
and therefore, for any $B$ and $\theta$, writing $\ket{\Psi_{0}(B,\theta,\phi)} \equiv \ket{\tilde{\Psi}_{0}(\phi)}$, we have
\be
\bra{\tilde{\Psi}_{0}(\phi)} \pd_{B} \ket{\tilde{\Psi}_{0}(\phi)}
=
\bra{\tilde{\Psi}_{0}(0)} \pd_{B} \ket{\tilde{\Psi}_{0}(0)}
=
0.
\ee
A similar reasoning holds for
$\bra{\Psi_{0}(B, \theta, \phi)} \pd_{\theta} \ket{\Psi_{0}(B, \theta, \phi)}$.
The only non-vanishing component is $A_{\phi}$, which reads
\beq
A_{\phi}
&=&
i \, \frac{1}{B \sin\theta} \, \bra{\tilde{\Psi}_{0}(\phi)} \pd_{\phi} \ket{\tilde{\Psi}_{0}(\phi)} 
\nn
\\
&=&
\frac{1}{B \sin\theta} \, \bra{\tilde{\Psi}_{0}(0)}\, i\,R(\phi) \pd_{\phi}R^{\dag}(\phi) \ket{\tilde{\Psi}_{0}(0)},
\nn
\\
\eeq
and since
\be
i\,R(\phi) \pd_{\phi} R^{\dag}(\phi)
=
i \, \e^{ i\, \phi \, \sigma_{\mathrm{tot}}^{z}/2 }  \, \pd_{\phi} \, \e^{ -i \, \phi \, \sigma_{\mathrm{tot}}^{z}/2 }
=
\frac{\sigma_{\mathrm{tot}}^{z}}{2},
\ee
the Berry connection in spherical coordinates is finally given by
\beq
\vec{A}^{(S)}(B,\theta,\phi)
&=&
\frac{1}{B \sin\theta} \, \bra{\tilde{\Psi}_{0}(0)} \frac{\sigma_{\mathrm{tot}}^{z}}{2}\ket{\tilde{\Psi}_{0}(0)} \, \hat{\phi}
\nn
\\
&=&
\frac{1}{B \sin\theta}\frac{\left\langle\sigma^z_{\mathrm{tot}}\right\rangle}{2} \hat{\phi},
\eeq
where 
$\left\langle\sigma^z_{\mathrm{tot}}\right\rangle$ is the ground-state
expectation value of the total magnetization in the $z-$direction at
$\phi=0$. 
In Cartesian coordinates, we have
\be
\vec{A}^{(C)}(x,y,z)
=
\frac{\left\langle\sigma^z_{\mathrm{tot}}\right\rangle}{2}\lp \frac{-y\,\hat{x} + x\,\hat{y}}{x^{2}+y^{2}}\rp,
\ee
with $\left\langle\sigma^z_{\mathrm{tot}}\right\rangle$, the ground-state
expectation value of the total magnetization in the $z-$direction
given in Cartesian coordinates at $B_{y}=0$.
The Berry curvature $\vec{F}$ is obtained by taking the curl of the
Berry connection.
\begin{figure*}[]
 \includegraphics[scale=0.5,trim=0.2mm 0mm 3mm 0.5mm, clip]{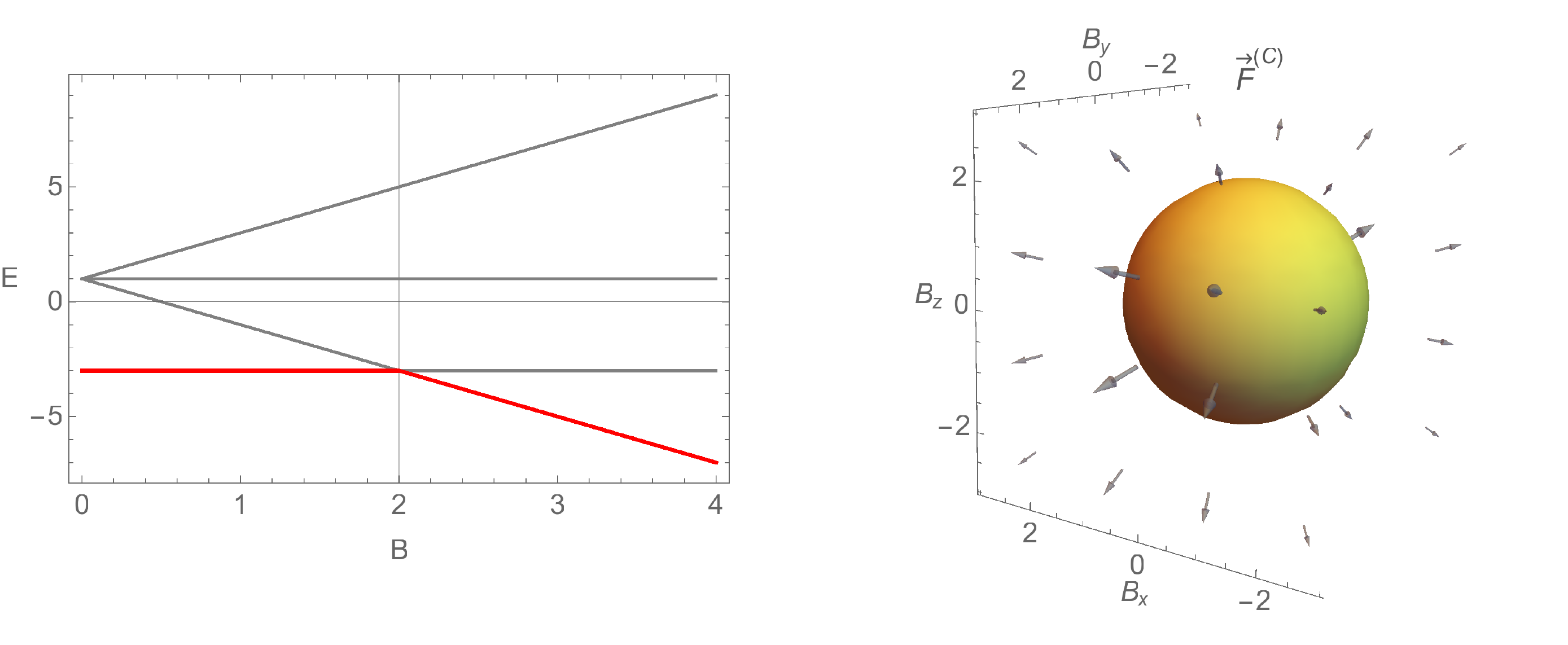}
\caption{\textit{Left panel:} The Energy spectrum of the Heisenberg
  Hamiltonian $H_{\mathrm{Heis}}$ as a function of $B$ is depicted. The
  ground-state energy $E_{0}(B)$ is shown by the thick red line.
  \textit{Right panel:} The Berry curvature in Cartesian
  coordinates~(\ref{eq:heisberc}) is plotted in parameter space
  $(B_{x},B_{y},B_{z})$.
  The sphere of radius $B=2$ carries a magnetic
  charge $q_{m}=2$, which is uniformly distributed over the surface of
  this sphere.
}
\label{fig:heisenbspf}
\end{figure*}

%%%%%%%%%%%%%%%%%%%%%%%%%%%%%%%%%%%%%%%%%%%%%%%%%%%%%%%%%%%%%%%%%%%%%%%%%%%%%%%%%%%%%%%%%%%%%%%%%%%%%%%%%%%%%%%%%%%%%%%%%%%%%%%%%%%%%%%%%%%%%%%%%%%%%%%%%%%%%%%%%%%%%%%%%%%%%%%%%%%%%%%%
\section{Heisenberg Interaction}
\label{app:B}
%%%%%%%%%

In this section, we set $\gamma_{1}=\gamma_{2}=g_{z}=1$, $g=2$ and
$B_{0}=0$ in the Hamiltonian~(\ref{eq:sysham}) in order to obtain the
two-qubit Heisenberg Hamiltonian
\be
H_{\mathrm{Heis}}
=
\vec{B}\cdot(\vec{\sigma}_{1}+\vec{\sigma}_{2})
+
\vec{\sigma}_{1}\cdot\vec{\sigma}_{2}.
\ee
This Hamiltonian has $SU(2)$ symmetry, which can be seen immediately
using spherical coordinates $(B,\theta,\phi)$
\be
H_{\mathrm{Heis}}(B, \theta, \phi)
=
B\,\hat{B}(\theta,\phi) \cdot (\vec{\sigma}_{1}+\vec{\sigma}_{2})
+
\vec{\sigma}_{1}\cdot\vec{\sigma}_{2}.
\ee
Namely, one observes that
\be
D(\hat{B}, \alpha) \, H_{\mathrm{Heis}} \, D^{\dag}(\hat{B}, \alpha)
=
H_{\mathrm{Heis}},
\ee
where
$D(\hat{B}, \alpha) = \exp[ i \, \alpha \, \hat{B}\cdot(\vec{\sigma}_{1}+\vec{\sigma}_{2}) ]$
is a generic element of $SU(2)$ and can be interpreted as a rotation
around the axis $\hat{B}(\theta,\phi)$ by an angle $\alpha$.
We note that we also have an exchange symmetry between the two qubits
$\vec{\sigma}_{1}\leftrightarrow\vec{\sigma}_{2}$.
As already mentioned previously, we can use the property
$H_{\mathrm{Heis}}(B, \theta, \phi) = R^\dag(\phi) H_{\mathrm{Heis}}(B, \theta, 0) R(\phi)$
to easily calculate the eigenenergies, eigenstates and thus the Berry
connection.
The ground-state energy is given by
\be
E_{0}(B)
=
\begin{cases} 
  - 3, & ~ B < 2, \\
  1 - 2 B, & ~ B > 2,
\end{cases}
\ee
and the corresponding ground-state reads
\beq
\ket{\tilde{\Psi}_{0}(\phi)}
=
\begin{cases}
  \frac{1}{\sqrt{2}} (0, 1, -1, 0)^{T},\\
  \lp e^{-i \phi} \sin^{2}\frac{\theta}{2},
   -\frac{\sin\theta}{2},
   -\frac{\sin\theta,}{2},
   \, e^{i\phi} \cos^{2}\frac{\theta}{2}\rp^{T}
\end{cases}
\nn \\
\eeq
for $B<2$ and $B>2$, respectively, where we used the basis
$\lbrace \ket{\uparrow\uparrow} = (1,0,0,0)^{T},
\, \ket{\uparrow\downarrow} = (0,1,0,0)^{T},
\, \ket{\downarrow\uparrow} = (0,0,1,0)^{T},
\, \ket{\downarrow\downarrow} = (0,0,0,1)^{T} \rbrace$, 
since
$\lbrace \ket{\uparrow}=(1,0)^{T}, \, \ket{\downarrow}=(0,1)^{T} \rbrace$
are the eigenstates of $\sigma_{i}^{z}$.
We observe that the ground-state degenerates on a sphere of radius
$B=2$ in the parameter space $\mathcal{M}\equiv \mathbb{R}^3$ defined in Cartesian coordinates by $(B_{x},B_{y},B_{z})$.
As illustrated in what follows, this degeneracy surface can be
interpreted as a magnetically charged sphere in parameter space
which creates an effective magnetic field, the Berry curvature $\vec{F}$.

The Berry connection in spherical coordinates can be calculated
explicitly, and we find
\be
\vec{A}^{(S)}(B,\theta,\phi)
=
A_{\phi} \, \hat{\phi}
=
 \begin{cases}
  0, & \quad B < 2, \\
  - \frac{1}{B} \cot\theta \, \hat{\phi} & \quad B > 2.
 \end{cases}
\ee
The Berry curvature, obtained by taking the curl of
$\vec{A}^{(S)}(B,\theta,\phi)$ in spherical coordinates, reads
\beq
\vec{F}^{(S)}(B,\theta,\phi)
&=&
\frac{1}{B \sin\theta}\, \pd_{\theta} \lp A_{\phi} \sin\theta \rp \hat{B},
\nn \\
\vec{F}^{(S)}(B,\theta,\phi)
&=&
\begin{cases}
  0, & \quad B < 2, \\
  \frac{1}{2}q_{m}\frac{1}{B^{2}}\hat{B}, & \quad B > 2,
\end{cases}
\label{eq:heisbercsp}
\eeq
where $q_{m}=2$ can be interpreted as an effective magnetic charge.
The Berry curvature allows us to read the first Chern number,
which indeed corresponds to the effective magnetic charge $q_m$,
\beq
\mathrm{ch}_{1}
&=&
\frac{1}{2\pi}
\iint_{\Sigma} \vec{F}^{(S)} \cdot d\vec{S},
\nn \\
\mathrm{ch}_{1}
&=&
\frac{1}{2\pi}
\int\limits_{0}^{\pi}\int\limits_{0}^{2\pi} \frac{1}{B^{2}}\, B^{2} \sin\theta\, d\theta\, d\phi
=
2
=
q_{m}.
\eeq
\begin{figure*}[]
 \includegraphics[scale=0.5,trim=0.2mm 0mm 3mm 0.5mm, clip]{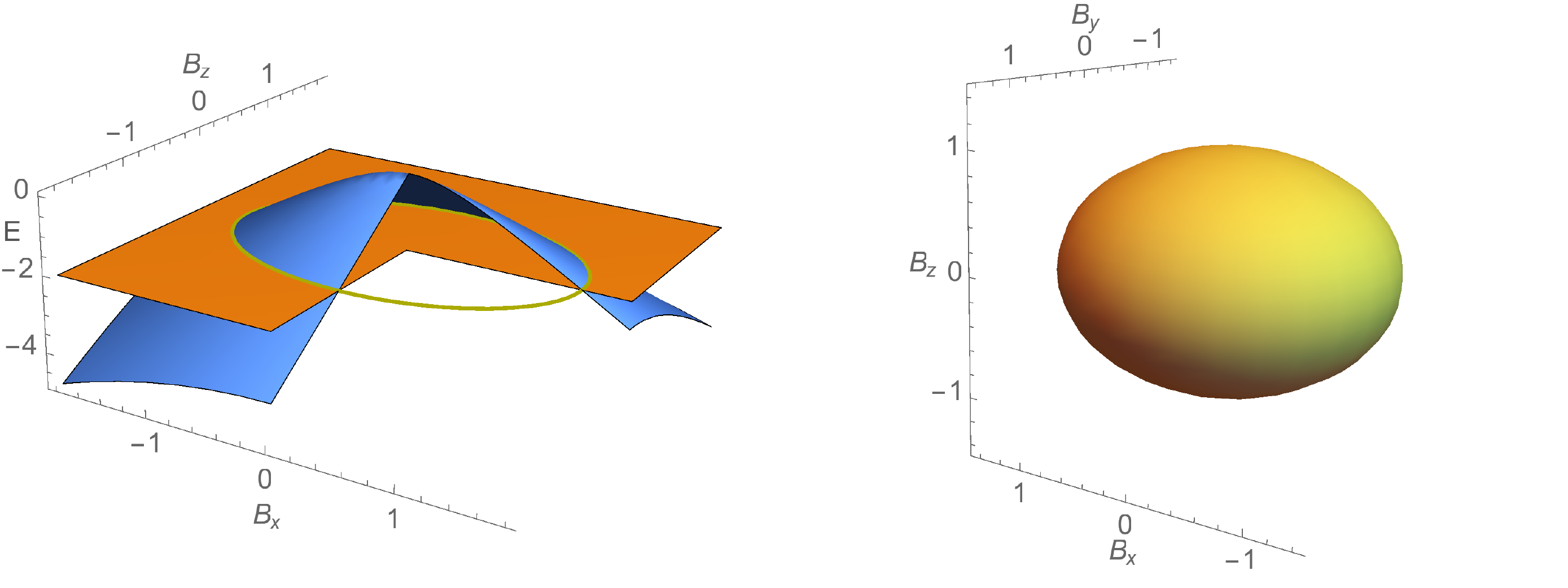}
\caption{\textit{Left panel:} The two lowest eigenenergies of
  $H_{\mathrm{XXZ}}$ are plotted as a function of $B_x$ and $B_z$ for
  $g_z=0.01$ and $B_{y}=0$.
  The singlet-state has constant energy given by $E_{\mathrm{singlet}}
  = -(2+g_z)$, and is shown by the constant (orange) plane, which cuts
  the (blue) surface corresponding to the eigenenergy of the next lowest
  eigenstate.
  The crossing curve, defined by the intersection of these two lowest
  eigenenergies, is given by an ellipse (yellow).
  \textit{Right panel:} The locus of the ground-state degeneracy
  defined by~(\ref{eq:locusellipsoids}), the surface of an
  ellipsoid, is plotted in parameter space $(B_{x},B_{y},B_{z})$. 
}
\label{fig:xxzcrossings}
\end{figure*}
\\
We used the fact that the surface element $d\vec{S}$ is strictly radial
$d\vec{S}=B^{2}\sin\theta\, d\theta \, d\phi\,\hat{B}$, and choose a spherical Gaussian surface $\Sigma$ centered at the origin with radius $B>2$ to calculate the above integral.
In Fig.~\ref{fig:heisenbspf} we show the spectrum of
$H_{\mathrm{Heis}}$ and the effective magnetic field given by the
Berry curvature
\be
\vec{F}^{(C)}(B_{x},B_{y},B_{z})
=
\begin{cases}
  0, \\
  \frac{1}{2}q_{m}\frac{\vec{B}}{(B_{x}^{2}+B_{y}^{2}+B_{z}^{2})^{3/2}},
\end{cases}
\label{eq:heisberc}
\ee
for  $\sqrt{B_{x}^{2}+B_{y}^{2}+B_{z}^{2}} < 2$ and $\sqrt{B_{x}^{2}+B_{y}^{2}+B_{z}^{2}} > 2$, respectively.
%
%
%\newpage
%%%%%%%%%%%%%%%%%%
\section{XXZ Interaction}
\label{app:C}

In this section, we choose $\gamma_{1} = \gamma_{2} = 1$, $g = 2$,
$B_{0} = 0$ and $g_{z} \neq 1$ in the Hamiltonian~(\ref{eq:sysham}),
which yields a two-qubit Hamiltonian with XXZ interaction
\begin{align}
H_{\mathrm{XXZ}}
&=
\vec{B} \cdot \left(\vec{\sigma}_{1} + \vec{\sigma}_{2} \right)
+
(\sigma_1^x \sigma_2^x + \sigma_1^y \sigma_2^y)
+
g_z \sigma_1^z\sigma_2^z
\nonumber \\
&=
\vec{B} \cdot \left(\vec{\sigma}_1 + \vec{\sigma}_2\right)
+
\vec{\sigma}_1 \cdot \vec{\sigma}_2
-
(1-g_z)\,\sigma_1^z\sigma_2^z.
\end{align}

This Hamiltonian is no longer $SU(2)$ symmetric but still has the
exchange symmetry between the two qubits.
Further, due to the property
$H_{\mathrm{XXZ}}(B, \theta, \phi) = R^\dag(\phi) H_{\mathrm{XXZ}}(B,
\theta, 0) R(\phi)$, we can set $B_{y}=0$ and using a more appropriate basis given by
$\{ \left\rvert\uparrow\uparrow\right\rangle,\,
    \left\rvert\downarrow\downarrow\right\rangle,\,
   (\left\rvert\uparrow\downarrow\right\rangle +
    \left\rvert\downarrow\uparrow\right\rangle)/\sqrt{2},\,
   (\left\rvert\uparrow\downarrow\right\rangle -
    \left\rvert\downarrow\uparrow\right\rangle)/\sqrt{2}\}$,
the Hamiltonian is written as a $4 \times 4$ matrix,
\beq
H_{\mathrm{XXZ}}
=
\begin{pmatrix}
2 B_z + g_z & 0 & \sqrt{2} \, B_x & 0 \\
0 & -2 B_z + g_z & \sqrt{2} \, B_x & 0 \\
\sqrt{2} \, B_x & \sqrt{2} \, B_x & 2 - g_z & 0 \\
0 & 0 & 0 & - 2 - g_z
\end{pmatrix}.
\nn
\\
\eeq
One can immediately see that the singlet-state
$(\left\rvert\uparrow\downarrow\right\rangle -
\left\rvert\downarrow\uparrow\right\rangle)/\sqrt{2}$ is an eigenstate
with eigenenergy $E_{\mathrm{singlet}}=-(2+g_{z})$.
More precisely, the singlet-state is the ground-state inside the locus
of crossing points given by 
\be
\frac{B_x^2}{2 \left(1 + g_z\right)} 
+
\frac{B_y^2}{2 \left(1 + g_z\right)}
+
\frac{B_z^2}{\left(1 + g_z\right)^2} = 1.
\label{eq:locusellipsoids}
\ee
The above expression defines the surface of an ellipsoid in the parameter
space $(B_{x},B_{y},B_{z})$, and can be obtained by solving the
equation for the energy crossing between the singlet-state and the
only other state with negative energy for $B_{y}=0$,
applying next the rotation
$R(\phi)=\exp(i\, \phi\, \sigma_{\mathrm{tot}}^{z}/2)$ to obtain the
result~(\ref{eq:locusellipsoids}) for $B_{y} \neq 0$ (see
Fig.~\ref{fig:xxzcrossings}).
The ground-state inside the ellipsoid is thus the Bell entangled
singlet-state
\be
\ket{\Psi_{0}}
=
\frac{1}{\sqrt{2}} (\ket{\uparrow\downarrow}-\ket{\uparrow\downarrow})
=
\frac{1}{\sqrt{2}} (0,1,-1,0)^{T},
\ee
and hence the Berry connection vanishes inside the ellipsoid, since
$\vec{A}^{(S)} = \frac{1}{B \sin\theta}
\left\langle\sigma^z_{\mathrm{tot}}\right\rangle \hat{\phi}\,$.
The Berry connection acquires only a non-zero value outside the
ellipsoid, which was calculated numerically using the standard
numerical diagonalization techniques.
The corresponding Berry curvature and its curl are depicted in
Fig.~\ref{fig:xxz-qm}.

%%%%%%

\subsection{Surface and Charge Density on the Ellipsoid}
\begin{figure*}[]
 \includegraphics[scale=0.7,trim=10.0mm 5mm 1mm 0mm, clip]{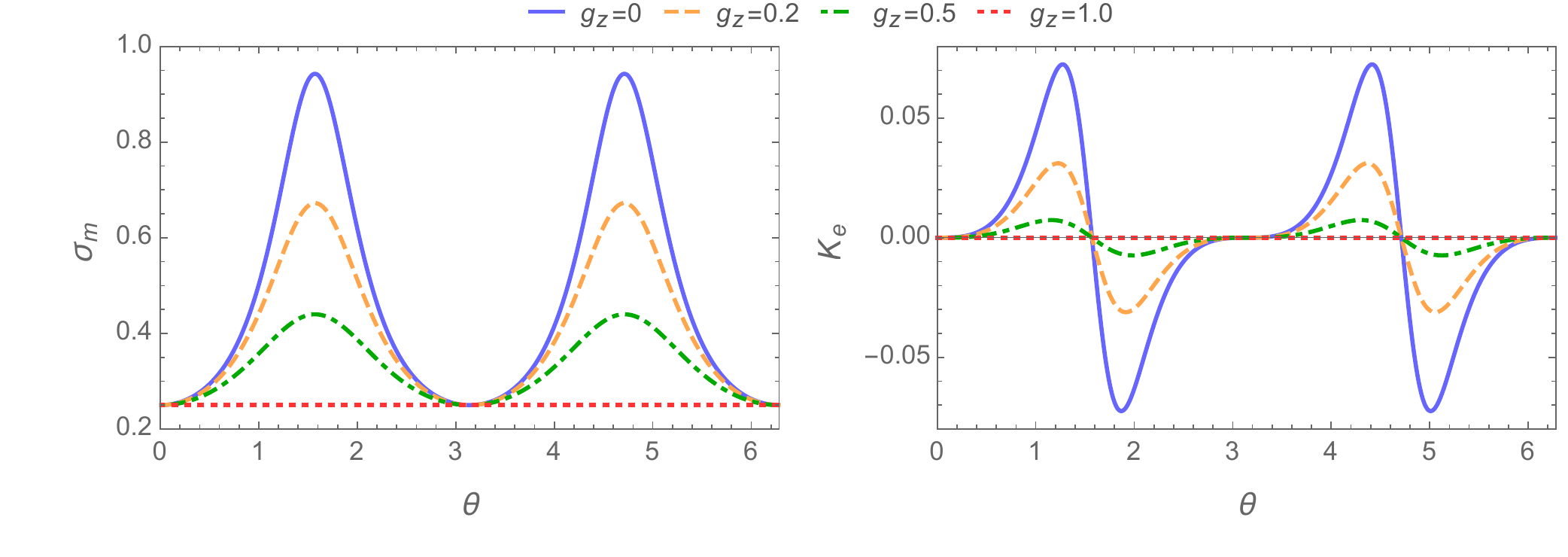}
 \caption{\textit{Left panel:} The surface charge density $\sigma_m$
   is shown as a function of $\theta$ for different values of $g_z$.
   Note that we plot $\theta$ from $0$ to $2\pi$, which means we go
   once around the entire ellipsoid. 
   One can see the charge accumulates on the equator as $g_z$
   decreases from $1$ to $0$.
   \textit{Right panel:} The surface current density $K_e$ as
   a function of $\theta$ for different $g_z$ values is depicted.
   $K_{e}$ changes sign at the equator ($\pi/2$
   and $3\pi/2$), indicating the quadrupole configuration pattern of
   $\vec{\nabla}\times\vec{F}$.}
 \label{fig:xxz-charge-current}
\end{figure*}
\begin{figure*}[]
 \includegraphics[scale=0.52,trim=4mm 0mm 0mm 0mm, clip]{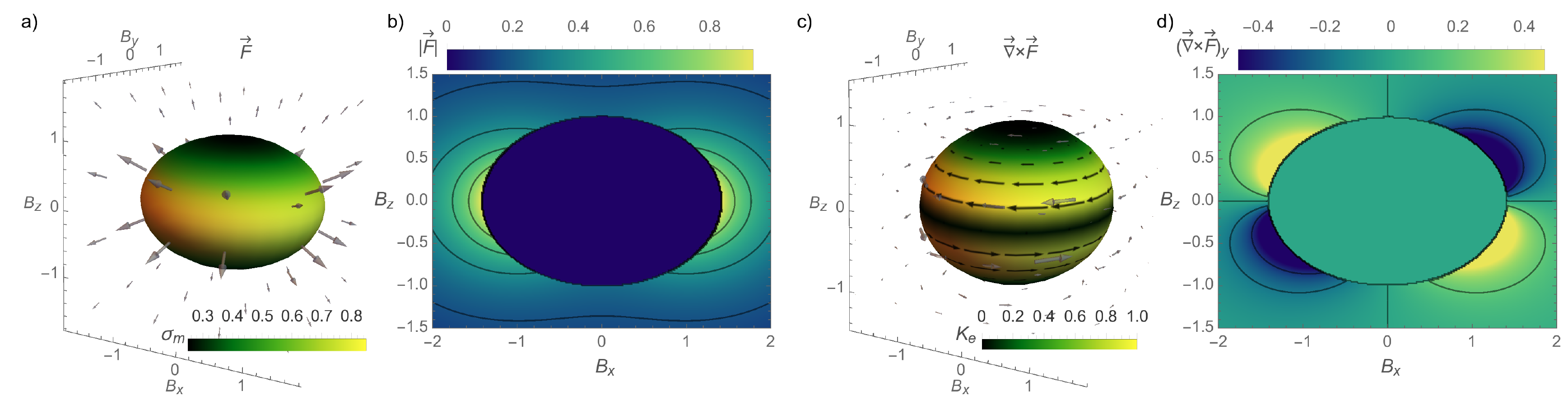}
 \caption{a) The surface charge density $\sigma_m$ on the
   ellipsoid with $\vec{F}$ for $g_{z}=0$. b) The corresponding
   $|\vec{F}|$ in the $xz$-plane. c) The surface electric current
   density $K_e$ with $\vec{\nabla}\times\vec{F}$ for $g_{z}=0$ and
   in d) we plotted $(\vec{\nabla}\times\vec{F})_{y}$ in the $xz$-plane.}
 \label{fig:xxz-qm}
\end{figure*}
The magnetic surface charge density $\sigma_m$ and effective electric
surface current $\vec{K}_e$ associated with the curl of $\vec{F}$ are
calculated in the following by considering the discontinuity in the
normal and parallel components of the magnetic field $\vec{F}$ across
the degeneracy surface (the ellipsoid)~\cite{Griffiths:611579}.
The magnetic surface charge density can be calculated from the
identity
\be
(F^{\perp}_\text{out} - F^{\perp}_\text{in}) = 2 \pi \sigma_m,
\ee
where $F^{\perp}_\text{out}$ ($F^{\perp}_\text{in}$) refers to the
perpendicular component of $\vec{F}$ just outside (inside) the
charged boundary surface.
The total charge $q_{m}$ in this example is obtained by
$
q_{m}
=
\frac{1}{2\pi}
\iint_{S} \, \sigma_{m}(\vec{r}) \, dS
$, where $dS$ is the differential area element of the ellipsoid surface.
Further, the electric surface current can be calculated through 
\be
\label{eq:surfacek}
\hat{n} \times (\vec{F}_\text{out} - \vec{F}_\text{in})
=
2\pi \vec{K}_e,
\ee
where $\vec{F}_\text{out}$ ($\vec{F}_\text{in}$) refers to $\vec{F}$
just outside (inside) the ellipsoid and $\hat{n}$ is a unit vector
perpendicular to the surface
Both the surface charge density and surface current density are
plotted in Fig.~\ref{fig:xxz-charge-current} versus the polar angle
$\theta$ for different values of $g_z$.
We note that the property
$H_{\mathrm{XXZ}}(B, \theta, \phi) = R^\dag(\phi) H_{\mathrm{XXZ}}(B, \theta, 0) R(\phi)$ 
implies that $\sigma_{m}$ and $\vec{K}_{e}$ are independent of the
azimuthal angle $\phi$.  
In Fig.~\ref{fig:xxz-qm} we plot $\sigma_{m}$ and $K_{e}$ on the
surface of the ellipsoid.
The total charge was also computed numerically
and found to be $q_{m}=+2$ for any value of $g_{z}$, as required
from topological considerations ($\mathrm{ch}_{1}=2$).
%
%%%%%%%%%%%%%%%%%%
\section{Anisotropic Interaction}
\label{app:D}

In this section, we set $\gamma_{1} = 1 + \alpha$,
$\gamma_{2} = 1-\alpha$, $g = 2$, $B_{0} = 0$ and $g_{z} =0$, with
$-1<\alpha<1$ in the Hamiltonian~(\ref{eq:sysham}), such that we
obtain a two-qubit Hamiltonian where the magnetic field $\vec{B}$
acts anisotropically on each spin
\be
H_{\mathrm{ani}}
=
\vec{B} \cdot
\left[ (1+\alpha) \, \vec{\sigma}_1 + (1-\alpha) \, \vec{\sigma}_2\right] 
+
\frac{g}{2} \, (\sigma_1^x\sigma_2^x
+
\sigma_1^y\sigma_2^y).
\ee
Writing this Hamiltonian in spherical coordinates, it can be seen that 
$H_{\mathrm{ani}}(B, \theta, \phi) = R^\dag(\phi) H_{\mathrm{ani}}(B, \theta, 0) R(\phi)$,
still holds
and therefore it is sufficient to study the spectrum for $B_{y}=0$.
Let us rewrite $H_{\mathrm{ani}}$ for $B_y = 0$ in the basis 
$\{ \left\rvert\uparrow\uparrow\right\rangle,\, \left\rvert\uparrow\downarrow\right\rangle, \, \left\rvert\downarrow\uparrow\right\rangle, \, \left\rvert\downarrow\downarrow\right\rangle \}$, defining $\alpha_{\pm} \equiv (1 \pm \alpha)$ for notational brevity,
\beq
H_{\mathrm{ani}}(B_{x}, B_{z})
=
\begin{pmatrix}
2 \, B_z & \alpha_- \, B_x & \alpha_+ \, B_x & 0 \\
\alpha_- \, B_x & 2 \,\alpha\, B_z & 2 & \alpha_+ \, B_x \\
\alpha_+ \, B_x & 2 & -2\,\alpha\, B_z & \alpha_- \,  B_x \\
0 & \alpha_+ \, B_x & \alpha_- \, B_x & -2\, B_z \\
\end{pmatrix}.
\nn \\
\eeq
\begin{figure*}[]
 \includegraphics[scale=0.5]{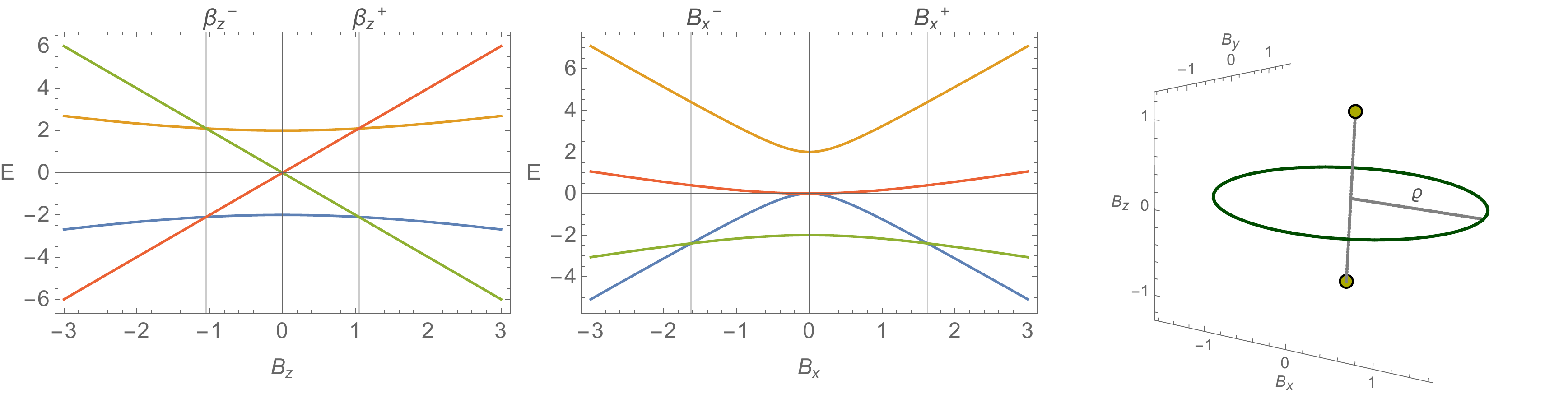}
 \caption{In the left panel we show a plot of the energy spectrum of
   $H_{\mathrm{ani}}(0, 0, B_{z})$, in the middle the eigenenergies of
   $H_{\mathrm{ani}}(B_{x}, 0, 0)$ and on the right panel we depict the
   ground-state degeneracies in the parameter space $(B_{x},B_{y},B_{z})$.}
 \label{fig:anisp}
\end{figure*}
\\
Obviously, for $B_x=0$, the matrix becomes block-diagonal, and the
ground-state energy crossings are located at 
\be
\beta_{z}^{(\pm)} = \pm \frac{1}{\sqrt{(1-\alpha^2)}}.
\ee
In the case of $B_{z}=0$, the Hamiltonian $H_{\mathrm{ani}}(B_{x},0,0)$
commutes with $\sigma_1^x\sigma_2^x$ and therefore they have a common
basis of eigenvectors given by
$
\lbrace
(-\ket{\uparrow\uparrow}+\ket{\downarrow\downarrow})/\sqrt{2},\,
(-\ket{\uparrow\downarrow}+\ket{\downarrow\uparrow})/\sqrt{2},\,
(\ket{\uparrow\uparrow}+\ket{\downarrow\downarrow})/\sqrt{2},\,
(\ket{\uparrow\downarrow}+\ket{\downarrow\uparrow})/\sqrt{2}\,
\rbrace
$.
With respect to this basis, $H_{\mathrm{ani}}(B_{x},0,0)$ is
block-diagonal,
\beq
H_{\mathrm{ani}}(B_{x}, 0, 0)
=
\begin{pmatrix}
0 & -2 \, \alpha \, B_x & 0 & 0 \\
-2 \, \alpha \, B_x & -2 & 0 & 0 \\
0 & 0 & 0 & 2 \, B_x \\
0 & 0 & 2 \, B_x & 2 \\
\end{pmatrix},
\nn \\
\eeq
where the corresponding eigenenergies can easily be calculated,
\beq
E_{1} = 1 - \sqrt{1+4\,B_{x}^{2}},
~
E_{2} &=& - 1 - \sqrt{1+4\,\alpha^{2} B_{x}^{2}},
\nn
\\
E_{3} = - 1 + \sqrt{1+4\,\alpha^{2} B_{x}^{2}},
~
E_{4} &=& 1 + \sqrt{1+4\,B_{x}^{2}},
\eeq
and it can be seen that the ground-state energy degenerates
($E_{1}=E_{2}$) at
$
B_{x}^{(\pm)} = \pm \sqrt{2\,(1+\alpha^2)}/(1-\alpha^2)
$.
The azimuthal invariance of the eigenenergies (the property
$H_{\mathrm{ani}}(B, \theta, \phi) = R^\dag(\phi) H_{\mathrm{ani}}(B,
\theta, 0) R(\phi)$), implies that these two points in the $x z-$plane actually correspond to a
ring centered at the origin in the $xy-$plane, with radius 
\be
\varrho =  \frac{\sqrt{2\,(1+\alpha^2)}}{(1-\alpha^2)}.
\ee
In Fig.~\ref{fig:anisp} we plot the spectrum of the two-qubit
Hamiltonian with an anisotropic magnetic field and the ground-state
degeneracies in parameter space $(B_{x},B_{y},B_{z})$, given by a ring
in the $xy$-plane and two points on the $z$-axis.

\begin{figure}[h]
 \includegraphics[scale=0.53]{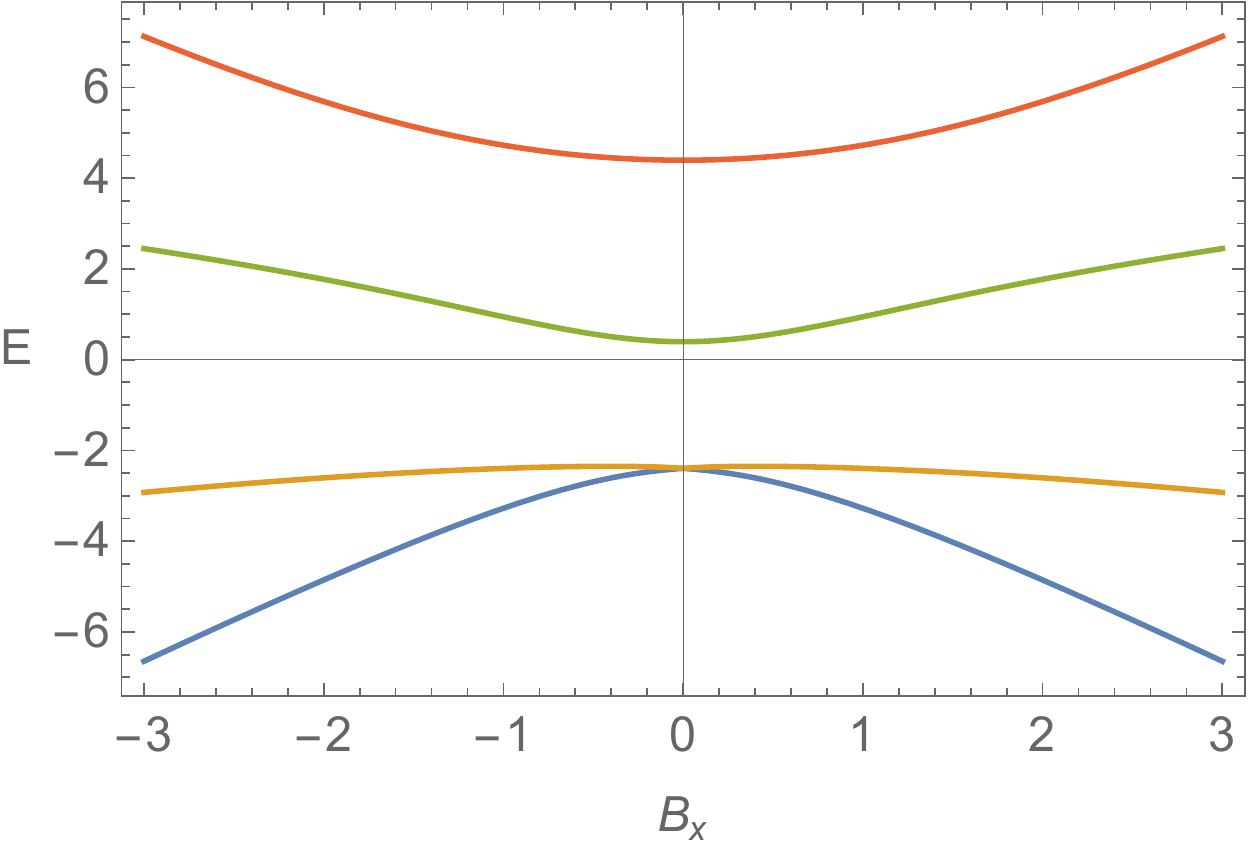}
 \caption{Renner-Teller level touching for $B_{x}=\varrho$ and
   changing $B_{z}$, i.e., vertically crossing the ring.}
 \label{fig:anileveltsp}
\end{figure}

Finally, we note that the ring has no charge, which can be seen
by calculating the first Chern number numerically,
$\mathrm{ch}_{1}(\mathrm{ring})=0$.
On the contrary, the two point charges on the $z$-axis carry each a
charge of $+1$. This was also confirmed by a numerical evaluation of
the first Chern number.
Energy-level crossings that yield a trivial Berry phase when encircled,
and therefore have an associated zero Chern number,
are know as Renner-Teller level touchings~\cite{Yarkony:1996wx}.
The energy level touching can be observed by fixing $B_{x}=\varrho$ and
varying $B_{z}$ (see Fig.~\ref{fig:anileveltsp}).

%%%%%%%%%%%%%%%%%%%%%%%%%%%%%%%%%%%%%%%%%%%%%%%%%%%%%%%%%%%%%%%%%%%%%%%%%%%%%%%%%
\section{Broken Exchange Symmetry}
\label{app:E}

In this section we calculate the Berry connection $\vec{A}$ (vector potential),
Berry curvature $\vec{F}$ (magnetic field) and the curl of the Berry
curvature $\nabla\times\vec{F}$ (current) using a degenerate
perturbation theory for the interacting two-qubit system with broken
exchange symmetry.
First, we derive the location of the ground-state degeneracies (level
crossings) in the parameter space.
Next, we compute the ground-state of the system up to
second-order using a degenerate perturbation theory.
The resulting ground-state allows us then to calculate the Berry
connection, Berry curvature and the curl of the Berry curvature in the
vicinity of the effective magnetic monopole charges (ground-state
degeneracies).

%%%%%%%%%%%%%%%%%%%%%%%%%%%%%%%%%%%%%%%%%%%%%%%%%%%%%%%%%%%%%%%%%%%%%%%%%%%%%%%%%%
\subsection{Location of the magnetic monopoles}

\begin{figure}[h]
  \begin{center}
    \includegraphics[scale=0.53,trim=1mm 1.5mm 2mm 1mm, clip]{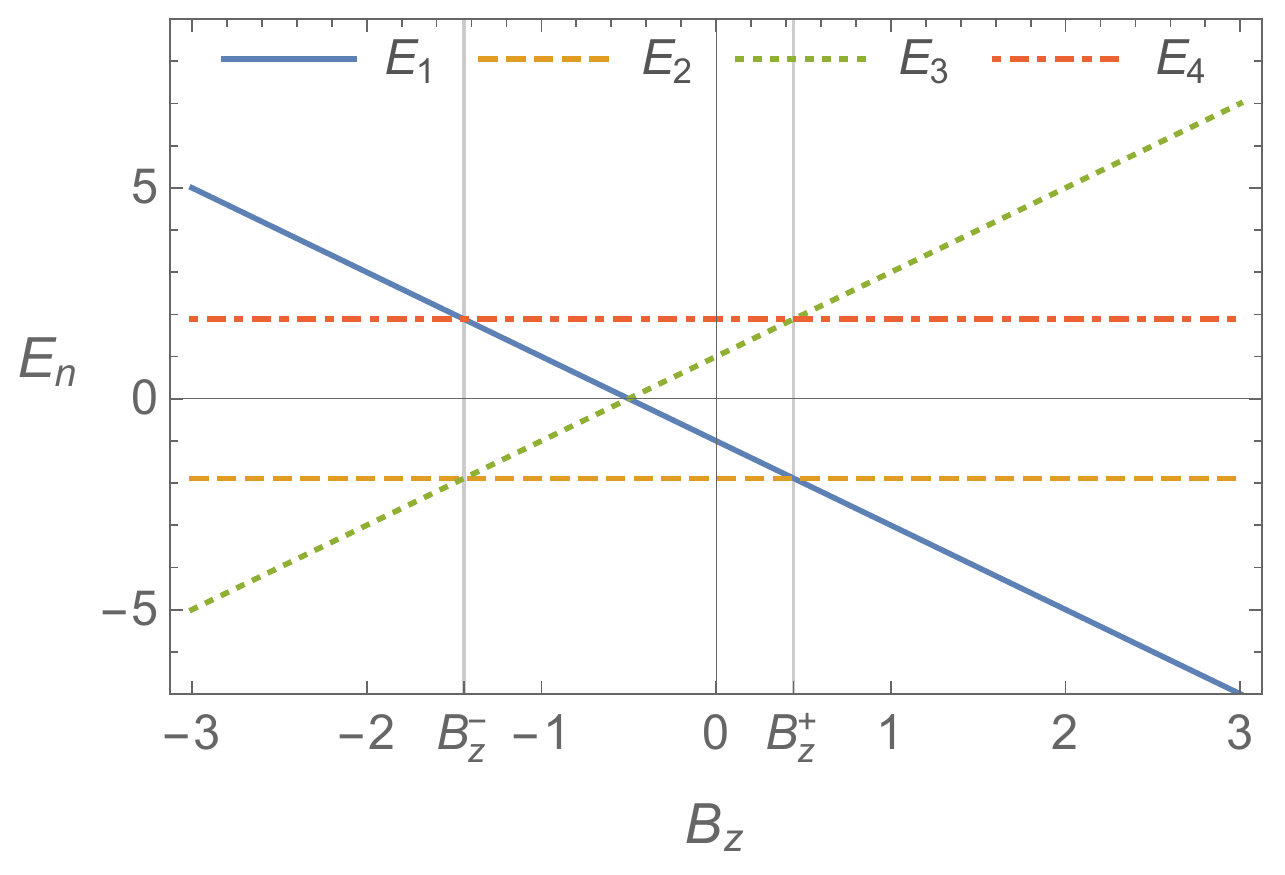}
    \caption{Eigenenergies as a function of $B_{z}$ for $B_{x}=B_{y}=0$
      and fixed $B_{0}$ and $g$. The degeneracies of the ground-state occur at $B_{z}=B_z^{-}$ and $B_{z}=B_z^{+}$.}
    \label{fig:eigvhbxby0}
  \end{center}
\end{figure}

The Hamiltonian for two interacting qubits with a broken exchange
symmetry is given by
\be
\label{eq:hamxybx}
H_{\mathrm{BES}}
=
\vec{B} \cdot (\vec{\sigma}_{1} + \vec{\sigma}_{2})
+
\frac{g}{2} (\sigma_{1}^{x} \sigma_{2}^{x} + \sigma_{1}^{y} \sigma_{2}^{y})
+
B_{0} \, \sigma_{1}^{z},
\ee
which can be obtained by setting $\gamma_{1} = \gamma_{2} = 1$ and
$g_{z} =0$ in the Hamiltonian~(\ref{eq:sysham}).
The ground-state degeneracies are restricted to the $B_{z}$
axis, since the eigenenergies of the Hamiltonian~(\ref{eq:hamxybx}) have
an azimuthal symmetry, and since the Hamiltonian~(\ref{eq:hamxybx}) itself has no more symmetries.
The positions of the ground-state degeneracies in parameter space can
therefore be determined by diagonalizing the
Hamiltonian~(\ref{eq:hamxybx}) for $B_{x}=B_{y}=0$.
In the basis $\{
\left\rvert\uparrow\uparrow\right\rangle,\,\left\rvert\uparrow\downarrow\right\rangle,\,\left\rvert\downarrow\uparrow\right\rangle,\,\left\rvert\downarrow\downarrow\right\rangle\}$
the Hamiltonian becomes block-diagonal
\beq
H_{\mathrm{BES}}(0,0,B_{z})
=
\begin{pmatrix}
B_{0} + 2 B_z & 0 & 0 & 0 \\
0 & B_0 & g & 0 \\
0 & g & -B_0 & 0 \\
0 & 0 & 0 & - B_{0} - 2 B_z \\
\end{pmatrix},
\nn \\
\eeq
and thus the corresponding eigenenergies $E_{n}$ and eigenstates
$\ket{\psi_{n}}$, with $n=1,2,3,4$, are
given by
\begin{align}
E_{1} &= -B_{0} - 2 B_{z},
\quad
\ket{\psi_{1}} = (0, 0, 0, 1)^{T},
\nonumber \\
E_{2} &= -\delta,
\quad
\ket{\psi_{2}} = \frac{1}{\sqrt{(B_z^{+})^{2} + \lp \frac{g}{2} \rp^{2}}} (0, -B_z^{+}, \frac{g}{2}, 0)^{T},
\nonumber \\
E_{3} &= B_{0} + 2 B_{z},
\qquad
\ket{\psi_{3}} = (1, 0, 0, 0)^{T},
\nonumber \\
E_{4} &= \delta,
\quad
\ket{\psi_{4}} = \frac{1}{\sqrt{(B_z^{-})^{2} + \lp \frac{g}{2} \rp^{2}}} (0, -B_z^{-}, \frac{g}{2}, 0)^{T},
\end{align}
where we defined
\beq
&\delta \equiv \sqrt{B_{0}^{2} + g^{2}},
\nonumber \\
&B_z^{+} \equiv \frac{1}{2}(-B_{0}+\delta),
\quad
B_z^{-} \equiv \frac{1}{2}(-B_{0}-\delta).
\eeq
In Fig.~\ref{fig:eigvhbxby0}, we plot the eigenenergies $E_{n}$ as a
function of $B_{z}$ for $B_{x}=B_{y}=0$ with fixed $B_{0}$ and $g$.
It shows that the ground-state energy-level crosses with the excited
energy-levels at $B_{z}=B_z^{-}$ and $B_{z}=B_z^{+}$.
These degeneracies act as magnetic monopoles in parameter space.
The ground-state energy of the system as a function of $B_{z}$ for
$B_{x}=B_{y}=0$ can be written as
\be
E_{0}(B_{z})
=
\begin{cases} 
  - B_{0} - 2 B_{z}, & ~ B_{z} \geq B_z^{+}, \\
  -\delta, & ~ B_z^{-} \leq B_{z} \leq B_z^{+}, \\
  B_{0} + 2 B_{z}, & ~ B_{z} \leq B_z^{-} \\
\end{cases}
\ee
and the corresponding ground-state reads
\be
\ket{\Psi_{0}(B_{z})}
=
\begin{cases}
  (0, 0, 0, 1)^{T}, & ~ B_{z} \geq B_z^{+}, \\
  \frac{(0, -B_z^{+},\, g/2,\, 0)^{T}}{\sqrt{(B_z^{+})^{2} \, + \,\lp g/2 \rp^{2}}}, & ~ B_z^{-} \leq B_{z} \leq B_z^{+}, \\
  (1, 0, 0, 0)^{T}, & ~ B_{z} \leq B_z^{-}. \\
\end{cases}
\ee

\section{Degenerate perturbation theory: Coordinate system centered at monopoles}
\label{app:F}

In what follows we use a degenerate perturbation theory to calculate
the ground-state of our two qubit system close to the degeneracies at
$B_z^{+}$ and $B_z^{-}$.
We will present only the results for small deviations around the
degeneracy $B_z^{+}$; the results around $B_z^{-}$ are
obtained in a similar way.
Let us consider the location of the degeneracy, given vectorially by
$\vec{B}'=(0,0,B_z^{+})^{T}$, as the origin of a new coordinate
system.
With respect to this new coordinate system, a point in parameter space
$(B_{x},B_{y},B_{z})$ is indicated by the vector
$d\vec{B}=(dB_{x},dB_{y},dB_{z})^{T}$ and it is related to the original
coordinate system by $d\vec{B}=\vec{B}-\vec{B}'$, which yields the following
relations between the two coordinate systems
\be
\label{eq:coordsystrelation}
\vec{B}
=
\begin{pmatrix}
B_{x} \\
B_{y} \\
B_{z}
\end{pmatrix}
=
\vec{B}' + d\vec{B}
=
\begin{pmatrix}
dB_{x} \\
dB_{y} \\
dB_{z} + B_{z}^+
\end{pmatrix}.
\ee
\begin{figure}[h]
  \begin{center}
    \includegraphics[scale=0.53]{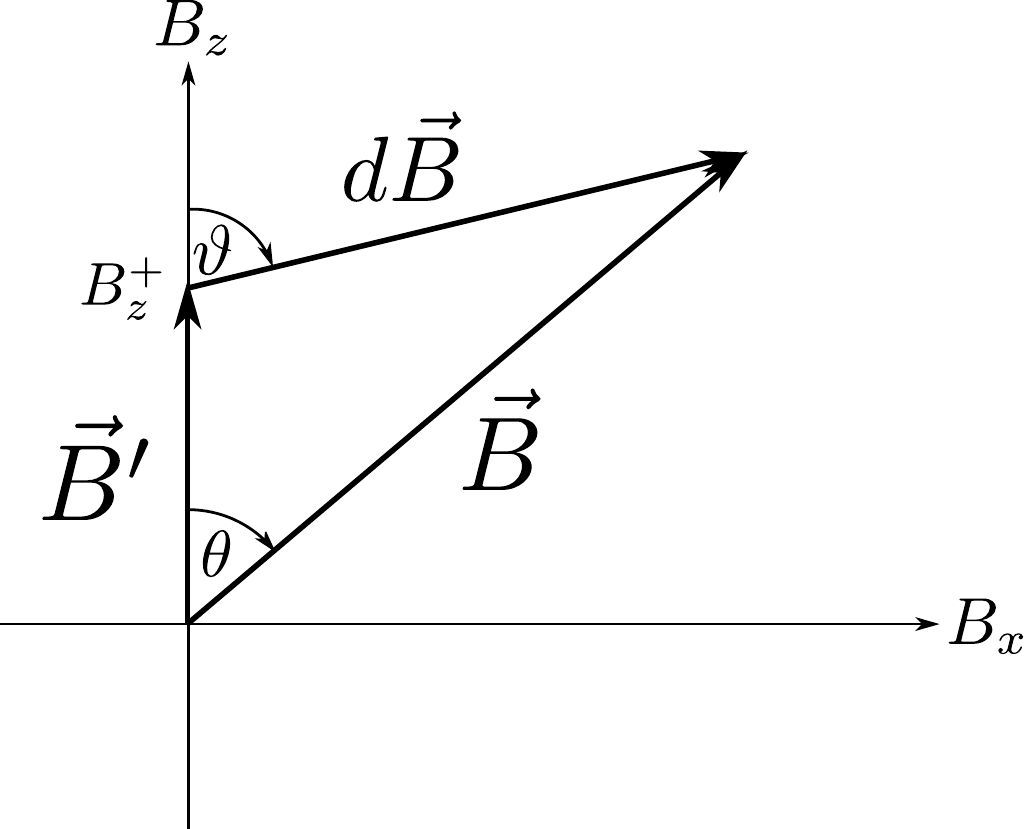}
    \caption{The two different coordinate systems: $\vec{B}$ and $d\vec{B}$.}
    \label{fig:coord}
  \end{center}
\end{figure}
\\
The Hamiltonian~(\ref{eq:hamxybx}) in the new coordinates reads
\beq
H_{\mathrm{BES}}
=
dB_{x} \, (\sigma_{1}^{x} + \sigma_{2}^{x})
+
dB_{y} \, (\sigma_{1}^{y} + \sigma_{2}^{y}) +
\nn \\
+
dB_{z} \, (\sigma_{1}^{z} + \sigma_{2}^{z})
+
B_{z}^+ (\sigma_{1}^{z} + \sigma_{2}^{z})
+
\nn \\
+
\frac{g}{2} (\sigma_{1}^{x} \sigma_{2}^{x} + \sigma_{1}^{y} \sigma_{2}^{y})
+
B_{0} \, \sigma_{1}^{z}.
\eeq
We treat the deviation from the monopole (degeneracy) as
a small perturbation, i.e., $|d\vec{B}| \ll 1$. It is therefore useful to express~(\ref{eq:coordsystrelation}) in spherical coordinates
\beq
d\vec{B}
=
\begin{pmatrix}
dB \sin\vartheta \cos\phi \\
dB \sin\vartheta \sin\phi \\
dB \cos\vartheta
\end{pmatrix}
=
\vec{B} - \vec{B}'
=
\begin{pmatrix}
B \sin\theta  \cos\phi \\
B \sin\theta \sin\phi \\
B \cos\theta - B_{z}^+
\end{pmatrix},
\nn \\
\eeq
which yields the relations
\be
dB \sin\vartheta = B \sin\theta,
~
dB \cos\vartheta = B \cos\theta - B_{z}^+.
\ee
Let us focus on the $xz-$plane defined by $B_{y}=0$, or in spherical
coordinates, by $\phi=0$.
Such choice implies $dB_{y}=0$, and thus we have
\begin{align}
H_{\mathrm{BES}}
&=
dB_{x} \, (\sigma_{1}^{x} + \sigma_{2}^{x})
+
dB_{z} \, (\sigma_{1}^{z} + \sigma_{2}^{z})
+
\nn \\
&\quad
+
B_{z}^+ (\sigma_{1}^{z} + \sigma_{2}^{z})
+
\frac{g}{2} (\sigma_{1}^{x} \sigma_{2}^{x} + \sigma_{1}^{y} \sigma_{2}^{y})
+
B_{0} \, \sigma_{1}^{z}
\nonumber
\\
&=
H_{0} + dB \, H'
\end{align}
where
\beq
H_{0}
&=&
B_{z}^+ (\sigma_{1}^{z} + \sigma_{2}^{z})
+
\frac{g}{2} (\sigma_{1}^{x} \sigma_{2}^{x} + \sigma_{1}^{y} \sigma_{2}^{y})
+
B_{0} \, \sigma_{1}^{z}, 
\nn \\
H'
&=&
\sin\vartheta \, (\sigma_{1}^{x} + \sigma_{2}^{x})
+
\cos\vartheta \, (\sigma_{1}^{z} + \sigma_{2}^{z})
.
\eeq

%%%%%%%%%%%%%%%%%%%%%%%%%%%%%%%%%%%%%%%%
\subsection{Degenerate perturbation theory: ground-state calculation}

First, we calculate the eigenvalues and eigenstates of
$H_{0}$.
The eigenenergies are given by
\beq
E_{1}^{(0)} = -\sqrt{B_{0}^{2}+g^{2}},
~
E_{2}^{(0)} &=& -\sqrt{B_{0}^{2}+g^{2}},
\nn \\
E_{3}^{(0)} = \sqrt{B_{0}^{2}+g^{2}},
~
E_{4}^{(0)} &=& \sqrt{B_{0}^{2}+g^{2}},
\eeq
and the corresponding eigenstates read
\begin{align}
\ket{\Psi_{1}^{(0)}}
&=
(0, 0, 0, 1)^{T},
\nn \\\
\ket{\Psi_{2}^{(0)}}
&=
(0, -\frac{B_{z}^+}{\sqrt{(B_{z}^+)^{2}+\lp\frac{g}{2}\rp^{2}}}, \frac{\lp\frac{g}{2}\rp}{\sqrt{(B_{z}^+)^{2}+\lp\frac{g}{2}\rp^{2}}}, 0)^{T},
\nonumber \\
\ket{\Psi_{3}^{(0)}}
&=
(0, -\frac{B_{z}^-}{\sqrt{(B_{z}^-)^{2}+\lp\frac{g}{2}\rp^{2}}}, \frac{\lp\frac{g}{2}\rp}{\sqrt{(B_{z}^-)^{2}+\lp\frac{g}{2}\rp^{2}}}, 0)^{T},
\nn \\
\ket{\Psi_{4}^{(0)}}
&=
(1, 0, 0, 0)^{T}.
\end{align}
The location of the energy level crossings of the ground-state and the
first excited state on the $z$-axis appear at
\beq
B_{z}^+ = \frac{1}{2}(-B_{0}+\sqrt{B_{0}^{2}+g^{2}}),
~
B_{z}^- = \frac{1}{2}(-B_{0}-\sqrt{B_{0}^{2}+g^{2}}).
\nn \\
\eeq
We note the following useful identities
\be
B_{z}^+ + B_{z}^- = - B_{0}, 
~
B_{z}^+ - B_{z}^- = \sqrt{B_{0}^{2}+g^{2}} \equiv \delta.
\ee
Let us also introduce the following notations,
\beq
\delta \equiv \sqrt{B_{0}^{2}+g^{2}}&,&
~
\beta^2 \equiv \frac{g}{\delta},
\nn \\
\Delta \equiv \frac{B_{z}^+-\frac{g}{2}}{\sqrt{(B_{z}^+)^{2}+\lp\frac{g}{2}\rp^{2}}}&,&
~
\eta
\equiv
\frac{(B_{z}^-)-\frac{g}{2}}{\sqrt{(B_{z}^-)^{2}+\lp\frac{g}{2}\rp^{2}}}.
\nn \\
\eeq
The unperturbed eigenstates
$\{ \ket{\Psi_{1}^{(0)}}, \ket{\Psi_{2}^{(0)}} \}$ and
$\{ \ket{\Psi_{3}^{(0)}}, \ket{\Psi_{4}^{(0)}} \}$
are degenerate, therefore one needs to use a degenerate perturbation theory to compute the first-order corrections.
To this end, we write the matrix
\beq
W
&=&
\begin{pmatrix}
\bra{\Psi_{1}^{(0)}} H' \ket{\Psi_{1}^{(0)}} & \bra{\Psi_{1}^{(0)}} H' \ket{\Psi_{2}^{(0)}} \\
\bra{\Psi_{2}^{(0)}} H' \ket{\Psi_{1}^{(0)}} & \bra{\Psi_{2}^{(0)}} H' \ket{\Psi_{2}^{(0)}}
\end{pmatrix}
\nn \\
&=&
\begin{pmatrix}
- 2 \cos\vartheta & - \frac{B_{z}^+-g/2 }{\sqrt{(B_{z}^+)^{2}+\lp g/2\rp^{2}}}\sin\vartheta \\
- \frac{B_{z}^+-g/2}{\sqrt{(B_{z}^+)^{2}+\lp g/2 \rp^{2}}}\sin\vartheta & 0
\end{pmatrix}
\nn \\
&=&
\begin{pmatrix}
- 2 \cos\vartheta & - \Delta \sin\vartheta \\
- \Delta \sin\vartheta & 0
\end{pmatrix}.
\eeq
The matrix $W$ has eigenvalues
\beq
E_{\pm}^{(1)}
&=&
- \cos\vartheta \pm \sqrt{\cos^{2}\vartheta + \Delta^{2}\sin^{2}\vartheta}
\nn \\
&=&
- \cos\vartheta \pm \sqrt{1-\beta^2 \sin^{2}\vartheta},
\eeq
where we used the identities
\beq
&&
\Delta^{2}
=
\frac{(B_{z}^+ - g/2)^{2}}{(B_{z}^+)^{2}+\lp g/2\rp^{2}}
=
1-\frac{2 B_{z}^+ \lp g/2\rp}{(B_{z}^+)^{2}+\lp g/2\rp^{2}}
\nn \\
&&
\qquad\qquad\qquad
\Delta^{2}
\equiv
1 - \frac{g}{\delta}
\equiv
1 - \beta^2 .
\eeq
The eigenvalues $E_{\pm}^{(1)}$ of the matrix $W$ give the first-order
correction to the two lowest eigenenergies, namely $E_{1,2} = E_{1}^{(0)} + dB\, E_{\pm}^{(1)}$.
The eigenvectors of $W$ are written as $\ket{w_{1,2}}$
and read 
\beq
&\ket{w_{1}}
=
\begin{pmatrix}
a_{1} \\
b_{1}
\end{pmatrix}
=
\begin{pmatrix}
\frac{E_{-}^{(1)}}{\sqrt{(E_{-}^{(1)})^{2} + \Delta^{2}\sin^{2}\vartheta}} \\
- \frac{\Delta \sin\vartheta}{\sqrt{(E_{-}^{(1)})^{2} + \Delta^{2}\sin^{2}\vartheta}}
\end{pmatrix},
\nn \\
&\ket{w_{2}}
=
\begin{pmatrix}
\frac{E_{+}^{(1)}}{\sqrt{(E_{+}^{(1)})^{2} + \Delta^{2}\sin^{2}\vartheta}} \\
- \frac{\Delta \sin\vartheta}{\sqrt{(E_{+}^{(1)})^{2} + \Delta^{2}\sin^{2}\vartheta}}
\end{pmatrix}.
\eeq
The ``good'' linear combination for the ground-state at zeroth-order
is therefore given by
\begin{widetext}
\begin{align}
\ket{\Psi_0^{(0)}}
&=
a_{1} \ket{\Psi_{1}^{(0)}} + b_{1} \ket{\Psi_{2}^{(0)}}
\nonumber \\
&=
\frac{E_{-}^{(1)}}{\sqrt{(E_{-}^{(1)})^{2} + \Delta^{2}\sin^{2}\vartheta}}
(0, 0, 0, 1)^{T}
+
\lp \frac{-\Delta \sin\vartheta}{\sqrt{(E_{+}^{(1)})^{2} + \Delta^{2}\sin^{2}\vartheta}}\rp
\lp 0, \frac{-B_{z}^+}{\sqrt{(B_{z}^+)^{2}+\lp\frac{g}{2}\rp^{2}}}, \frac{\lp\frac{g}{2}\rp}{\sqrt{(B_{z}^+)^{2}+\lp\frac{g}{2}\rp^{2}}}, 0 \rp^{T}
\nonumber \\
&=
\frac{1}{\sqrt{(E_{-}^{(1)})^{2} + \Delta^{2}\sin^{2}\vartheta}}
\lp 0, \frac{B_{z}^+ \Delta \sin\vartheta}{\sqrt{(B_{z}^+)^{2}+\lp\frac{g}{2}\rp^{2}}}, \frac{-\lp\frac{g}{2}\rp \Delta \sin\vartheta}{\sqrt{(B_{z}^+)^{2}+\lp\frac{g}{2}\rp^{2}}}, E_{-}^{(1)}\rp^{T}.
\end{align}
\end{widetext}
The first-order correction to the ground-state is
\begin{widetext}
\begin{align}
\ket{\Psi_0^{(1)}}
&=
\sum_{n \neq \{1,2\}}
\frac{\bra{\Psi_{n}^{(0)}} H' \ket{\Psi_0^{(0)}}}{( E_{1}^{(0)}-E_{n}^{(0)})} \ket{\Psi_{n}^{(0)}}
=
-\frac{1}{2\,\delta}
\frac{1}{\sqrt{(E_{-}^{(1)})^{2} + \Delta^{2}\sin^{2}\vartheta}}
\begin{pmatrix}
\Delta^{2} \sin^{2}\vartheta \\
\frac{B_z^-}{B_z^--\frac{g}{2}} \, \eta^{2} \, E_{-}^{(1)} \, \sin\vartheta \\
-\frac{\frac{g}{2}}{B_z^--\frac{g}{2}} \, \eta^{2} \, E_{-}^{(1)} \, \sin\vartheta \\
0
\end{pmatrix}.
\end{align}
\end{widetext}
The second-order correction to the ground-state is given by
\begin{widetext}
\begin{align}
\ket{\Psi_0^{(2)}}
&=
\sum_{k \neq \{1,2\}} \sum_{l \neq \{1,2\}}
\frac{\bra{\Psi_{k}^{(0)}} H' \ket{\Psi_{l}^{(0)}}\bra{\Psi_{l}^{(0)}} H' \ket{\Psi_0^{(0)}}}{(E_{1}^{(0)}-E_{k}^{(0)})(E_{1}^{(0)}-E_{l}^{(0)})}
\ket{\Psi_{k}^{(0)}}
-
\frac{1}{2} \ket{\Psi_0^{(0)}} 
\sum_{k \neq \{1,2\}}
\frac{\bra{\Psi_0^{(0)}} H' \ket{\Psi_{k}^{(0)}}\bra{\Psi_{k}^{(0)}} H' \ket{\Psi_0^{(0)}}}{(E_{1}^{(0)}-E_{k}^{(0)})^{2}}
+
\nonumber \\
&-
\sum_{k \neq \{1,2\}}
\frac{\bra{\Psi_0^{(0)}} H' \ket{\Psi_0^{(0)}}\bra{\Psi_{k}^{(0)}} H' \ket{\Psi_0^{(0)}}}{(E_{1}^{(0)}-E_{k}^{(0)})^{2}}
\ket{\Psi_{k}^{(0)}}
\\
&=
\frac{1}{4\, \delta^{2}}
\frac{\Delta^{2}\sin^{2}\vartheta}{\sqrt{(E_{-}^{(1)})^{2} + \Delta^{2}\sin^{2}\vartheta}}
\begin{pmatrix}
\frac{\eta^{2}}{\Delta^{2}} E_{-}^{(1)} + 2 \cos\vartheta \\
\frac{B_z^-}{B_z^--\frac{g}{2}} \, \eta^{2} \sin\vartheta \\
-\frac{\frac{g}{2}}{B_z^--\frac{g}{2}} \, \eta^{2} \sin\vartheta \\
0
\end{pmatrix}
+
\nonumber \\
&-
\frac{1}{8\, \delta^{2}}
\frac{\Delta^{2}\sin^{2}\vartheta}{\sqrt{(E_{-}^{(1)})^{2} + \Delta^{2}\sin^{2}\vartheta}}
\frac{(\frac{\eta^{2}}{\Delta^{2}}(E_{-}^{(1)})^{2} + \Delta^{2}\sin^{2}\vartheta)}{(E_{-}^{(1)})^{2} + \Delta^{2}\sin^{2}\vartheta}
\begin{pmatrix}
0 \\
\frac{B_{z}^+}{B_{z}^+-\frac{g}{2}} \, \Delta^{2} \sin\vartheta \\
-\frac{\frac{g}{2}}{B_{z}^+-\frac{g}{2}} \, \Delta^{2} \sin\vartheta \\
E_{-}^{(1)}
\end{pmatrix}
+
\nonumber \\
&+
\frac{1}{2\, \delta^{2}}
\frac{E_{-}^{(1)}\sin\vartheta}{\sqrt{(E_{-}^{(1)})^{2} + \Delta^{2}\sin^{2}\vartheta}}
\frac{(E_{-}^{(1)} \cos\vartheta - \Delta^{2}\sin^{2}\vartheta)}{(E_{-}^{(1)})^{2} + \Delta^{2}\sin^{2}\vartheta}
\begin{pmatrix}
\Delta^{2} \sin\vartheta \\
\frac{B_z^-}{B_z^--\frac{g}{2}} \, \eta^{2} \, E_{-}^{(1)} \\
-\frac{\frac{g}{2}}{B_z^--\frac{g}{2}} \, \eta^{2} \, E_{-}^{(1)} \\
0
\end{pmatrix}.
\end{align}
\end{widetext}
The ground-state can then be expressed up to second-order by
\be
\ket{\Psi_0 (dB,\vartheta,0)}
=
\ket{\Psi_0^{(0)}} + dB\, \ket{\Psi_0^{(1)}} + dB^{2} \, \ket{\Psi_0^{(2)}},
\ee
and the dependence on the azimuth angle $\phi$ is obtained through the
following rotation
\be
\ket{\Psi_0 (dB,\vartheta,\phi)} = R^{\dag}(\phi)\, \ket{\Psi_0 (dB,\vartheta,0)},
\label{eq:rotprops}
\ee
where
$R(\phi) = \exp(i\, \phi\, ( \op{\sigma}_{1}^{z} + \op{\sigma}_{2}^{z} ) /2)$.
%

%%%%%%%%%%%%%%%%%%%%%%%%%%%%%%%%%
\subsection{Berry connection $-$ Effective magnetic vector potential}

We are now able to calculate the Berry connection in spherical
coordinates $\vec{A}_{+}^{(S)}(dB,\vartheta,\phi)$, where the $+$ sign
indicates that we are considering small radial deviations $dB$ close to the degeneracy located at $B_{z}^{+}$.
The operator $\vec{\nabla}$ in spherical coordinates
$(dB, \vartheta, \phi)$ is given by
\be
\vec{\nabla}
=
\left( 
\frac{\pd}{\pd(dB)},\, \frac{1}{dB}\, \frac{\pd}{\pd\vartheta},\, \frac{1}{dB \sin\vartheta}\,\frac{\pd}{\pd\phi} \right)^{T},
\ee
therefore the only non zero component of the Berry connection is
$A_{\phi,+}$ and reads
\begin{widetext}
\begin{align}
A_{\phi,+}
&=
i\,\frac{1}{dB \sin\vartheta} \bra{\Psi_0 (dB,\vartheta,\phi)} \pd_{\phi} \ket{\Psi_0 (dB,\vartheta,\phi)} 
\nonumber \\
&=
\frac{1}{dB \sin\vartheta}
\bra{\Psi_0 (dB,\vartheta,0)}
\frac{1}{2}
\lp \op{\sigma}_{1}^{z} + \op{\sigma}_{2}^{z} \rp
\ket{\Psi_0 (dB,\vartheta,0)}
\nonumber \\
&=
\frac{1}{dB \sin\vartheta}
\Bigg[
\bra{\Psi_0^{(0)}} \frac{1}{2} \lp \op{\sigma}_{1}^{z} + \op{\sigma}_{2}^{z} \rp \ket{\Psi_0^{(0)}}
+
2 \, dB \, \bra{\Psi_0^{(0)}} \frac{1}{2} \lp \op{\sigma}_{1}^{z} + \op{\sigma}_{2}^{z} \rp \ket{\Psi_0^{(1)}}
~+
\nonumber \\
&~~~~~~~~~~~~~~+
dB^{2}
\lp
 2 \, \bra{\Psi_0^{(0)}} \frac{1}{2} \lp \op{\sigma}_{1}^{z} + \op{\sigma}_{2}^{z} \rp \ket{\Psi_0^{(2)}}
 +
 \bra{\Psi_0^{(1)}} \frac{1}{2} \lp \op{\sigma}_{1}^{z} + \op{\sigma}_{2}^{z} \rp \ket{\Psi_0^{(1)}}
\rp
+ \ldots
\Bigg]
\nonumber \\
&=
\frac{1}{dB \sin\vartheta}
\Bigg[
-\frac{(E_{-}^{(1)})^{2}}{(E_{-}^{(1)})^{2} + \Delta^{2}\sin^{2}\vartheta}
+
dB^{2}
\frac{1}{4\,\delta^{2}}
\frac{\Delta^{4}\sin^{4}\vartheta}{(E_{-}^{(1)})^{2} + \Delta^{2}\sin^{2}\vartheta}
~+
\nonumber \\
&\phantom{\frac{1}{b \sin\vartheta}\Bigg[}~~~~+
dB^{2}
\frac{1}{4\,\delta^{2}}
\lp
\frac{\eta^{2}}{\Delta^{2}}(E_{-}^{(1)})^{2} + \Delta^{2}\sin^{2}\vartheta
\rp
\frac{(E_{-}^{(1)})^{2}}{(E_{-}^{(1)})^{2} + \Delta^{2}\sin^{2}\vartheta}
\frac{\Delta^{2}\sin^{2}\vartheta}{(E_{-}^{(1)})^{2} + \Delta^{2}\sin^{2}\vartheta}
+ \ldots
\Bigg]
\nonumber \\
&=
\frac{1}{dB \sin\vartheta}
\Bigg[
-\frac{1}{2}
\lp 1 + \frac{\cos\vartheta}{\sqrt{1-\beta^2 \sin^{2}\vartheta}} \rp
+
dB^{2} \frac{1}{8\,\delta^{2}}
\lp 1 - \frac{\cos\vartheta}{\sqrt{1-\beta^2 \sin^{2}\vartheta}} \rp
(1-\beta^2) \sin^{2}\vartheta
~+
\nonumber \\
&\phantom{\frac{1}{b \sin\vartheta}\Bigg[}~~~~+
dB^{2} \frac{1}{8\,\delta^{2}}
\sin^{2}\vartheta
\lp
1
+
\frac{(1+\beta^2)\cos\vartheta}{\sqrt{1-\beta^2 \sin^{2}\vartheta}}
+
\frac{\beta^2 \cos^{2}\vartheta}{1-\beta^2 \sin^{2}\vartheta}
\rp
+ \ldots
\Bigg]
\nonumber \\
&\approx
\frac{1}{dB \sin\vartheta}
\Bigg[
-\frac{1}{2}
\lp 1 + \frac{\cos\vartheta}{\sqrt{1-\beta^2 \sin^{2}\vartheta}} \rp
+
dB^{2} \frac{1}{8\,\delta^{2}} \sin^{2}\vartheta
\lp
2-\beta^2
+
\frac{2\beta^2\cos\vartheta}{\sqrt{1-\beta^2 \sin^{2}\vartheta}}
+
\frac{\beta^2 \cos^{2}\vartheta}{1-\beta^2 \sin^{2}\vartheta}
\rp
\Bigg],
\label{eq:aphipt}
\end{align}
\end{widetext}
where 
\be
\beta^2
=
\frac{g}{\delta}
=
\frac{g}{\sqrt{B_{0}^{2}+g^{2}}},
\qquad
\frac{\eta^{2}}{\Delta^{2}} = \frac{1+\beta^2}{1-\beta^2}.
\ee
In summary, one can write the Berry connection (vector potential)
in spherical coordinates 
$\vec{A}_{+}^{(S)}(dB, \vartheta, \phi)
=
A_{\phi,+} \, \hat{\phi}
$,
with
\begin{widetext}
\begin{align}
A_{\phi,+}
&\approx
\frac{1}{dB \sin\vartheta}
\Bigg[
-\frac{1}{2}
\lp 1 + \frac{\cos\vartheta}{\sqrt{1-\beta^2 \sin^{2}\vartheta}} \rp
+
dB^{2}\, \frac{1}{8\, \delta^{2}} \sin^{2}\vartheta
\lp
2-\beta^2
+
\frac{2\beta^2\cos\vartheta}{\sqrt{1-\beta^2 \sin^{2}\vartheta}}
+
\frac{\beta^2 \cos^{2}\vartheta}{1-\beta^2 \sin^{2}\vartheta}
\rp
\Bigg],
\end{align}
\end{widetext}
and keeping only the leading order term for $A_{\phi,+}$ we have
\be
\label{eq:aphi-leading}
A_{\phi,+} \approx - \frac{1}{2}\, \frac{1}{ dB \sin\vartheta} \left( 1 + \frac{\cos\vartheta}{\sqrt{1 - \beta^2 \sin^2 \vartheta}} \right).
\ee
\begin{figure*}[]
\begin{tabular}{cc}
 \includegraphics[scale=0.6]{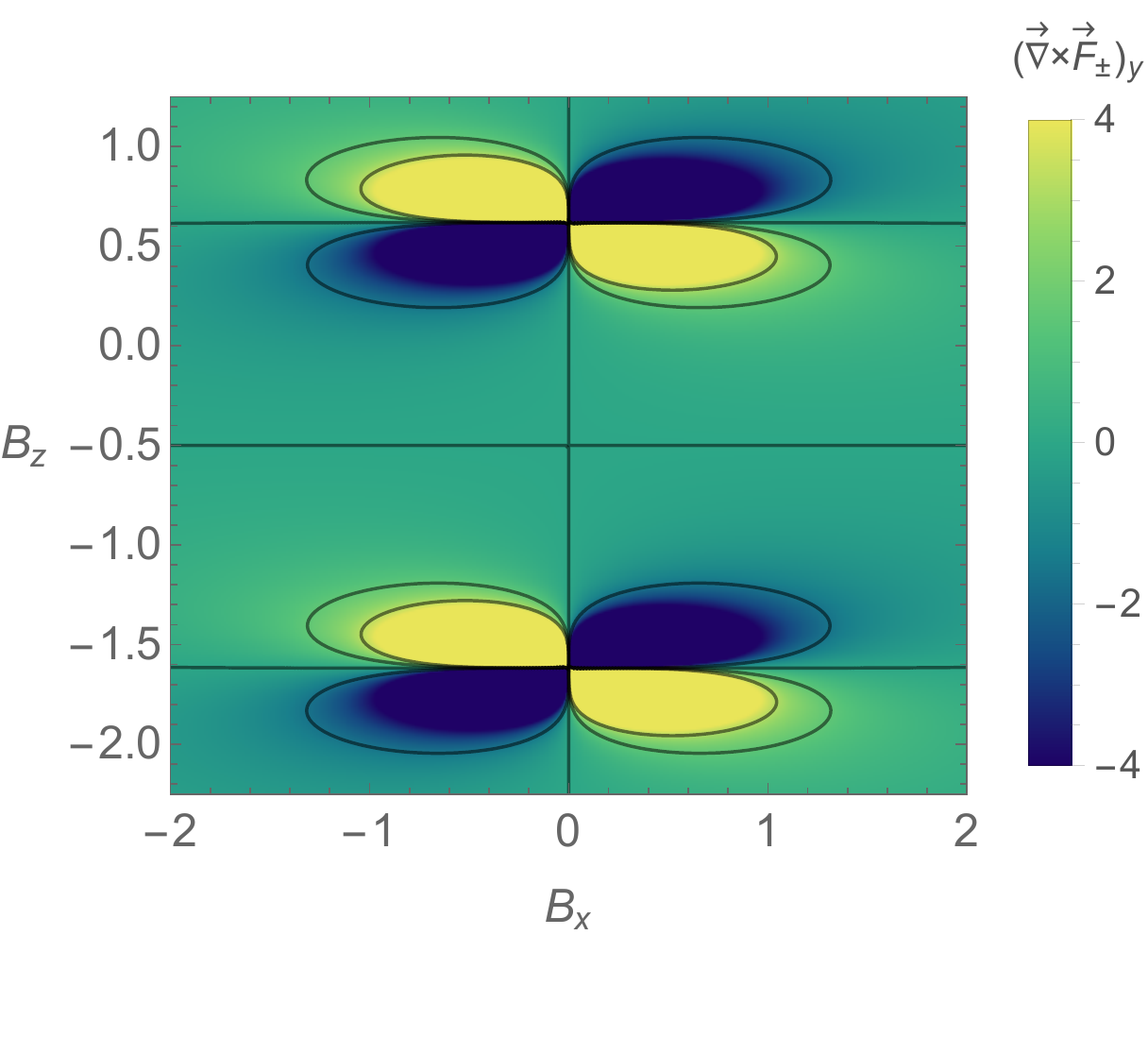}
 &
 \includegraphics[scale=0.6]{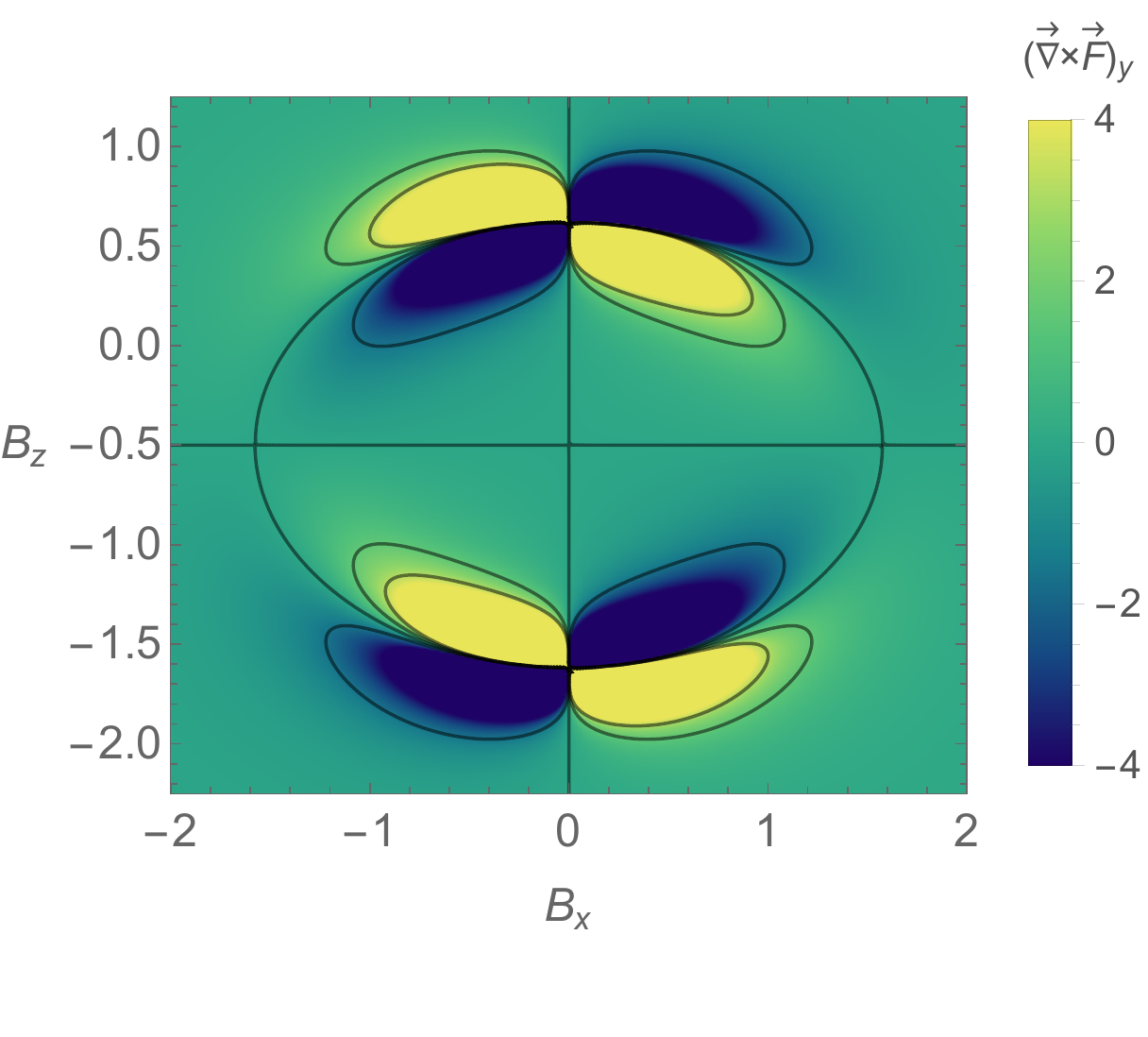}
\end{tabular}
 \caption{A density plot of the $y$ component of
   $\vec{\nabla}\times\vec{F}$ in Cartesian coordinates as
   a function of $B_{x}$ and $B_{z}$ is shown for $B_{y}=0$, $g=2$ and
   $B_{0}=1$.
   The curl of the Berry curvature has only a $y$ component in the
   plane defined by $B_{y}=0$.
   Negative values indicate that the vectors point perpendicularly out of the plane
   and positive values indicate the vectors point perpendicularly into the plane.
   On the left panel we show the curl of the Berry curvature obtained
   by perturbation theory and on the right obtained by exact
   diagonalization.}
 \label{fig:pertcurlfsp}
\end{figure*}

%%%%%%%%%%%%%%%%%%%%%%%%%%%%%%%%%%%%%%%%%%%%%%%%%%%%%%%%%%%%%%%%%%%%%%%%%%%%%%%%%
\subsection{Berry curvature $-$ Effective magnetic field}

The Berry curvature $\vec{F}^{(S)}(dB,\vartheta,\phi)$ 
is obtained by taking the curl of~(\ref{eq:aphi-leading}).
The curl operator in spherical coordinates $(dB,\vartheta,\phi)$ reads
\begin{widetext}
\begin{align}
&\vec{F}^{(S)}(dB,\vartheta,\phi) = \vec{\nabla} \times \vec{A}^{(S)}(dB,\vartheta,\phi) =
\nonumber \\
&=
\frac{1}{dB \sin\vartheta}
\left(
 \pd_{\vartheta} \left( A_{\phi} \sin\vartheta \right)
 -
 \pd_{\phi} A_{\vartheta}
\right)
\hat{dB}
+
\frac{1}{dB}
\left(
 \frac{1}{\sin\vartheta} \pd_{\phi} A_{dB}
 -
 \pd_{dB} \left(dB A_{\phi} \right)
\right)
\hat{\vartheta}
+
\frac{1}{dB}
\left(
 \pd_{dB} \left( dB A_{\vartheta} \right)
 -
 \pd_{\vartheta} A_{dB}
\right)
\hat{\phi},
\end{align}
\end{widetext}
where
$\vec{A}^{(S)}(dB,\vartheta,\phi) = A_{dB}\hat{dB} + A_{\vartheta}\hat{\vartheta} + A_{\phi}\hat{\phi}$.
The only non vanishing component of 
$\vec{A}_{+}^{(S)}(dB, \vartheta, \phi)$ is $A_{\phi,+}$ and hence in
the leading order of $dB$ we find
\beq
\vec{F}_{+}^{(S)}(dB,\vartheta,\phi)
&=&
\vec{\nabla} \times\vec{A}_{+}^{(S)}(dB,\vartheta,\phi)
\nn \\
&\approx&
\frac{1}{2} \, \frac{1}{\gamma^2 \,(1 - \beta^2 \sin^2\vartheta )^{3/2}}
\frac{1}{dB^2} \hat{dB},
\nn \\
\eeq
where we introduced $\gamma \equiv 1/\sqrt{1-\beta^2}$.

\subsection{Curl of Berry curvature}

Finally, the curl of the Berry curvature to the leading order in $dB$
near the monopole $B_{z}^{+}$ can be calculated, and reads
\be
\label{eq:pertfcurlsup}
\vec{\nabla} \times \vec{F}_{+}^{(S)}(dB,\vartheta,\phi)
\approx
- \frac{3}{4}
\,
\frac{\beta^2 \sin 2\vartheta}{\gamma^2\,(1-\beta^2\sin^2\vartheta)^{5/2}}
\frac{1}{dB^3} \, \hat{\phi}~.
\ee
Following exactly the same procedure described above, but applied
to the degeneracy located at $B_{z}^{-}$, one finds that the leading
order of $A_{\phi,-}$ is given by
\be
\label{eq:aphi-leading-neg}
A_{\phi,-}
\approx
\frac{1}{2}\, \frac{1}{ dB \sin\vartheta} \left( 1 - \frac{\cos\vartheta}{\sqrt{1 - \beta^2 \sin^2 \vartheta}} \right),
\ee
with respect to the coordinate system centered on $B_{z}^{-}$.
The Berry curvature $\vec{F}_{-}^{(S)}(dB,\vartheta,\phi)$ and  
$\vec{\nabla} \times \vec{F}_{-}^{(S)}(dB,\vartheta,\phi)$ can then be
calculated accordingly.
The curl of the Berry curvature with respect to the original Cartesian
coordinate system $(B_{x},B_{y},B_{z})$ takes then the form
\begin{widetext}
\be
\vec{\nabla} \times \vec{F}_{(\pm)}^{(C)}
\approx
-
\frac{3}{2}
\frac{\beta^{2} B_{x} (B_{z}-B_{z}^{(\pm)})}
     {\gamma^{2} \lb (1-\beta^{2}) B_{x}^{2} + (B_{z}-B_{z}^{(\pm)})^{2} \rb^{5/2}}
\lp\frac{-B_{y} \, \hat{x} + B_{x} \, \hat{y}}{\sqrt{B_{x}^{2} + B_{y}^{2}}}\rp.
\ee
\end{widetext}
In the $B_{x}-B_{z}$ plane, corresponding to $B_{y}=0$, only the
$y$-component of $\vec{\nabla} \times \vec{F}_{(\pm)}^{(C)}$ is non-zero.
This situation is plotted in Fig.~\ref{fig:pertcurlfsp}.
For comparison we also plot the $y$-component of the curl of the Berry
curvature calculated numerically by using exact diagonalization.

\end{appendix}

% Create the reference section using BibTeX:
%%%%%%%%%%%%\bibliography{BibTex.bib}

% ****** End of file apstemplate.tex ******

\end{document}